\documentclass[11pt]{article}
\usepackage{xypic}
\usepackage{mathrsfs}
\usepackage{amssymb}
\usepackage{amsfonts}
\usepackage{amsmath}

\def\ben{\begin{enumerate}}
\def\een{\end{enumerate}}

\def\bit{\begin{itemize}}
\def\eit{\end{itemize}}

\def\0{\leqno}

\input xy
\xyoption{all}
\input{tcilatex}
\begin{document}

\begin{center}
{\Large {LEGENDRE DUALITY BETWEEN }\ \ \bigskip }

{\Large {\ LAGRANGIAN AND HAMILTONIAN MECHANICS}\bigskip }


\textbf{by }

\textbf{CONSTANTIN M. ARCU\c{S} }
\end{center}

\bigskip

\bigskip

\bigskip



%

\ \

\begin{abstract}
In some previous papers, a Legendre duality between Lagrangian and
Hamiltonian Mechanics has been developed. The $\left( \rho ,\eta \right) $%
-tangent application of the Legendre bundle morphism associated to a
Lagrangian L or Hamiltonian H is presented. Using that, a Legendre
description of Lagrangian Mechanics and Hamiltonian Mechanics is developed.
Duality between Lie algebroids structure, adapted $\left( \rho ,\eta \right)
$-basis, distinguished linear $\left( \rho ,\eta \right) $-connections and
mechanical $\left( \rho ,\eta \right) $-systems is the scope of this paper.
In the particular case of Lie algebroids, new results are presented. In the
particular case of the usual Lie algebroid tangent bundle, the classical
results are obtained. \ \ \bigskip\newline
\textbf{2000 Mathematics Subject Classification:} 00A69, 58A15, 53B05,
53B40, 58B34, 71B01, 53C05, 53C15, 53C60, 58E15, 70G45, 70H20, 70S05.\bigskip%
\newline
\ \ \ \textbf{Keywords:} fiber bundle, vector bundle, (generalized) Lie
algebroid, (linear) connection, curve, lift, natural base, adapted base,
projector, almost product structure, almost tangent structure, complex
structure, spray, semispray, mechanical system, Lagrangian formalism,
Hamiltonian formalism.
\end{abstract}

\tableofcontents

\section{Introduction}

The notion of Lagrange space was introduced and studied by J. Kern $\left[ 14%
\right] $ and R. Miron $\left[ 20\right] .$ In the Classical Mechanics we
obtain a Lagrangian formalism if we use $TTM$ as space for developing the
theory. (see $\left[ 6,7,21,22,26,27,28,34\right] $)

In $[18]$, P. Liberman showed that such a Lagrangian formalism is not
possible if the tangent bundle of a Lie algebroid is considered as space for
developing the theory. In his paper $[39]$ A. Weinstein developed a
generalized theory of Lagrangian Mechanics on Lie algebroids and obtained
the equations of motion, using the linear Poisson structure on the dual of
the Lie algebroid and the Legendre transformation associated with the
regular Lagrangian $L.$ In that paper, he asks the question of whether it is
possible to develop a formalism on Lie algebroids similar to Klein's
formalism $\left[ 16\right] $ in ordinary Lagrangian Mechanics. This task
was finally done by E. Martinez in $[19]$ (see also $[9,18]$).

The geometry of algebroids was extensively studied by many authors. (see$%
\left[ 3,29,30,\right] $, $\left[ 35,36,37,38\right] $)

Using the generalized Lie algebroids, (see $\left[ 1,2\right] $) a new point
of view over the Lagrangian Mechanics is presented in the paper $\left[ 4%
\right] $. The canonical $\left( \rho ,\eta \right) $-(semi)spray associated
to mechanical $\left( \rho ,\eta \right) $-system $\left( \left( E,\pi
,M\right) ,F_{e},\left( \rho ,\eta \right) \Gamma \right) $\ and from
locally invertible $\mathbf{B}^{\mathbf{v}}$-morphism $\left( g,h\right) $
have been presented. Lagrange mechanical $\left( \rho ,\eta \right) $%
-systems, the spaces necessary to solve the Weinstein's problem\textbf{, }%
were introduced. There have been presented the $\left( \rho ,\eta \right) $%
-semispray associated to a regular Lagrangian $L$ and external force $F_{e}$
which are applied on the total space of a generalized Lie algebroid and the
equations of Euler-Lagrange type.

The geometry of $T^{\ast }M$ is from one point of view different from that
of $TM,$ because it does not exist a natural tangent structure and a
semispray can not be introduced in the case of the tangent bundle. Two
geometrical ingredients are of great importance to $T^{\ast }M$: the
canonical $1$-form $p_{i}dx^{i}$ and its exterior derivative $dp_{i}\wedge
dx^{i}$ (the canonical symplectic structure of $T^{\ast }M$). They are
systematically used to define new useful tools in the classical theory.

The concept of Hamilton space, introduced in $\left[ 25\right] $ and
intensively studied in $\left[ 10,11,\right] $ $\left[ 12,13,21,22\right] $
has been succesful, as a geometric theory of the Hamiltonian function. The
modern formulation of the geometry of Cartan spaces was given by R. Miron $%
\left[ 23,24\right] $ although some results were obtained by \'{E}. Cartan $%
\left[ 8\right] $ and A. Kawaguchi $\left[ 15\right] .$

A Hamiltonian description of Mechanics on duals of Lie algebroids was
presented in $\left[ 17\right] .$ (see also $[32,33,35,36,37,38$) The role
of cotangent bundle of the configuration manifold is played by the
prolongation $\mathcal{L}^{\tau ^{\ast }}E$ of $E$ along the projection $%
\begin{array}[b]{ccc}
E^{\ast } & ^{\underrightarrow{~\tau ^{\ast }\ }} & M.%
\end{array}%
$ The Lie algebroid version of the classical results concerning the
universality of the standard Liouville $1$-form on cotangent bundles is
presented. Given a Hamiltonian function $%
\begin{array}[b]{ccc}
E^{\ast } & ^{\underrightarrow{~H\ }} & \mathbb{R}%
\end{array}%
$ and the symplectic form $\Omega _{E}$ on $E^{\ast },$ the dynamics are
obtained solving the equation%
\begin{equation*}
i_{\xi _{H}}\Omega _{E}=d^{\mathcal{L}^{\tau ^{\ast }}E}H
\end{equation*}%
with the usual notations. The solutions of $\xi _{H}$ (curves in $E^{\ast }$%
) are the ones of the Hamilton equations for $H.$ (see $\left[ 17\right] $)

A Hamiltonian description of Mechanics on duals of generalized Lie
algebroids without the symplectic form is presented in the paper $\left[ 5%
\right] .$ The canonical $\left( \rho ,\eta \right) $-semispray associated
to the dual mechanical $\left( \rho ,\eta \right) $-system $\left( \left(
\overset{\ast }{E},\overset{\ast }{\pi },M\right) ,\overset{\ast }{F}%
_{e},\left( \rho ,\eta \right) \Gamma \right) $\ and from locally invertible%
\emph{\ }$\mathbf{B}^{\mathbf{v}}$-morphism $\left( g,h\right) .$ Also, the
canonical $\left( \rho ,\eta \right) $-spray associated to mechanical system
$\left( \left( \overset{\ast }{E},\overset{\ast }{\pi },M\right) ,\overset{%
\ast }{F}_{e},\left( \rho ,\eta \right) \Gamma \right) $\ and from locally
invertible $\mathbf{B}^{\mathbf{v}}$-morphism\emph{\ }$(g,h)$ have been
presented$.$

The Hamilton mechanical $\left( \rho ,\eta \right) $-systems are the spaces
necessary to obtain a Hamiltonian formalism\textbf{\ }in the general
framewrok of generalized Lie algebroids. The $\left( \rho ,\eta \right) $%
-semispray associated to a regular Hamiltonian $H$ and external force $%
\overset{\ast }{F}_{e}$ which are applied on the dual of the total space of
a generalized Lie algebroid and the equations of Hamilton-Jacobi type have
been presented.

Using the classical Legendre transformation different geometrical objects on
$TM$ are naturally related to similar ones on $T^{\ast }M.$ The geometry of
Hamilton space can be obtained from that of a certain Lagrange space and
vice versa. As a particular case, we can associate its dual to a given
Finsler which is a Cartan space. In addition, in some conditions the $L$%
-dual of Kropina space is the Randers space and the $L$-dual of Randers
space is the Kropina space. These spaces are used in several applications in
Physics. (see $\left[ 13\right] $)

The Legendre transformation $%
\begin{array}[b]{ccc}
E & ^{\underrightarrow{~Leg_{L}\ }} & E^{\ast }%
\end{array}%
$ associated with a Lagrangian $L$ induces a Lie algebroid morphism $%
\begin{array}[b]{ccc}
\mathcal{L}^{\tau }E & ^{\underrightarrow{~\mathcal{L}Leg_{L}\ }} & \mathcal{%
L}^{\tau ^{\ast }}E,%
\end{array}%
$ which permits in the regular case to connect Lagrangian and Hamiltonian
formalisms as in Classical Mechanics. (see $\left[ 17\right] $)

In this paper we extend our study in the general framework of generalized
Lie algebroids. Using the $\left( \rho ,\eta \right) $-tangent application
of the Legendre bundle morphism associated to a Lagrangian $L$ or
Hamiltonian $H$ we obtain a lot of new results.

The Lagrangian Mechanics presented in the paper $\left[ 4\right] $ is dual
to the Hamiltonian Mechanics presented in the paper $\left[ 5\right] $ and
vice versa. In paricular, a new point of view over the Legendre duality
between Lagrangian Mechanics and Hamiltonian Mechanics in the framework of
Lie algebroids is presented in this paper.

\section{Preliminaries}

Let$\mathbf{~Vect},$ $\mathbf{Liealg},~\mathbf{Mod}$\textbf{,} $\mathbf{%
Man,~B}$ and $\mathbf{B}^{\mathbf{v}}$ be the category of real vector
spaces, Lie algebras, modules, manifolds, fiber bundles and vector bundles
respectively.

We know that if $\left( E,\pi ,M\right) \in \left\vert \mathbf{B}^{\mathbf{v}%
}\right\vert $ so that $M$ is paracompact and if $A\subseteq M$ is closed,
then for any section $u$ over $A$ it exists $\tilde{u}\in $ $\Gamma \left(
E,\pi ,M\right) $ so that $\tilde{u}_{|A}=u.$ In the following, we consider
only vector bundles with paracompact base.

Aditionally, if $\left( E,\pi ,M\right) \in \left\vert \mathbf{B}^{\mathbf{v}%
}\right\vert ,$ $\Gamma \left( E,\pi ,M\right) =\left\{ u\in \mathbf{Man}%
\left( M,E\right) :u\circ \pi =Id_{M}\right\} $ and $\mathcal{F}\left(
M\right) =\mathbf{Man}\left( M,\mathbb{R}\right) ,$ then $\left( \Gamma
\left( E,\pi ,M\right) ,+,\cdot \right) $ is a $\mathcal{F}\left( M\right) $%
-module. If \ $\left( \varphi ,\varphi _{0}\right) \in \mathbf{B}^{\mathbf{v}%
}\left( \left( E,\pi ,M\right) ,\left( E^{\prime },\pi ^{\prime },M^{\prime
}\right) \right) $ such that $\varphi _{0}\in Iso_{\mathbf{Man}}\left(
M,M^{\prime }\right) ,$ then, using the operation
\begin{equation*}
\begin{array}{ccc}
\mathcal{F}\left( M\right) \times \Gamma \left( E^{\prime },\pi ^{\prime
},M^{\prime }\right) & ^{\underrightarrow{~\ \ \cdot ~\ \ }} & \Gamma \left(
E^{\prime },\pi ^{\prime },M^{\prime }\right) \\
\left( f,u^{\prime }\right) & \longmapsto & f\circ \varphi _{0}^{-1}\cdot
u^{\prime }%
\end{array}%
\end{equation*}%
it results that $\left( \Gamma \left( E^{\prime },\pi ^{\prime },M^{\prime
}\right) ,+,\cdot \right) $ is a $\mathcal{F}\left( M\right) $-module and we
obtain the $\mathbf{Mod}$-morphism%
\begin{equation*}
\begin{array}{ccc}
\Gamma \left( E,\pi ,M\right) & ^{\underrightarrow{~\ \ \Gamma \left(
\varphi ,\varphi _{0}\right) ~\ \ }} & \Gamma \left( E^{\prime },\pi
^{\prime },M^{\prime }\right) \\
u & \longmapsto & \Gamma \left( \varphi ,\varphi _{0}\right) u%
\end{array}%
\end{equation*}%
defined by
\begin{equation*}
\begin{array}{c}
\Gamma \left( \varphi ,\varphi _{0}\right) u\left( y\right) =\varphi \left(
u_{\varphi _{0}^{-1}\left( y\right) }\right) ,%
\end{array}%
\end{equation*}%
for any $y\in M^{\prime }.$

Let $M,N\in \left\vert \mathbf{Man}\right\vert ,$ $h\in Iso_{\mathbf{Man}%
}\left( M,N\right) $ and $\eta \in Iso_{\mathbf{Man}}\left( N,M\right) $ be.

We know (see $\left[ 1,2\right] $) that if $\left( F,\nu ,N\right) \in
\left\vert \mathbf{B}^{\mathbf{v}}\right\vert $ so that there exists
\begin{equation*}
\begin{array}{c}
\left( \rho ,\eta \right) \in \mathbf{B}^{\mathbf{v}}\left( \left( F,\nu
,N\right) ,\left( TM,\tau _{M},M\right) \right)%
\end{array}%
\end{equation*}%
and also an operation
\begin{equation*}
\begin{array}{ccc}
\Gamma \left( F,\nu ,N\right) \times \Gamma \left( F,\nu ,N\right) & ^{%
\underrightarrow{\left[ ,\right] _{F,h}}} & \Gamma \left( F,\nu ,N\right) \\
\left( u,v\right) & \longmapsto & \left[ u,v\right] _{F,h}%
\end{array}%
\end{equation*}%
with the following properties:\bigskip

\noindent $\qquad GLA_{1}$. the equality holds good
\begin{equation*}
\begin{array}{c}
\left[ u,f\cdot v\right] _{F,h}=f\left[ u,v\right] _{F,h}+\Gamma \left(
Th\circ \rho ,h\circ \eta \right) \left( u\right) f\cdot v,%
\end{array}%
\end{equation*}%
\qquad \quad\ \ for all $u,v\in \Gamma \left( F,\nu ,N\right) $ and $f\in
\mathcal{F}\left( N\right) .$

\medskip $GLA_{2}$. the $4$-tuple $\left( \Gamma \left( F,\nu ,N\right)
,+,\cdot ,\left[ ,\right] _{F,h}\right) $ is a Lie $\mathcal{F}\left(
N\right) $-algebra,

$GLA_{3}$. the $\mathbf{Mod}$-morphism $\Gamma \left( Th\circ \rho ,h\circ
\eta \right) $ is a $\mathbf{LieAlg}$-morphism of
\begin{equation*}
\left( \Gamma \left( F,\nu ,N\right) ,+,\cdot ,\left[ ,\right] _{F,h}\right)
\end{equation*}%
source and
\begin{equation*}
\left( \Gamma \left( TN,\tau _{N},N\right) ,+,\cdot ,\left[ ,\right]
_{TN}\right)
\end{equation*}%
target, \medskip \noindent then the triple $\left( \left( F,\nu ,N\right) ,%
\left[ ,\right] _{F,h},\left( \rho ,\eta \right) \right) $ is called
generalized Lie algebroid.\emph{\ }

In particular, if $h=Id_{M}=\eta ,$ then we obtain the definition of the Lie
algebroid.

Let $\left( \left( F,\nu ,N\right) ,\left[ ,\right] _{F,h},\left( \rho ,\eta
\right) \right) $ be ageneralized Lie algebroid.

\begin{itemize}
\item Locally, for any $\alpha ,\beta \in \overline{1,p},$ we set $\left[
t_{\alpha },t_{\beta }\right] _{F,h}=L_{\alpha \beta }^{\gamma }t_{\gamma }.$
We easily obtain that $L_{\alpha \beta }^{\gamma }=-L_{\beta \alpha
}^{\gamma },~$for any $\alpha ,\beta ,\gamma \in \overline{1,p}.$
\end{itemize}

The real local functions $L_{\alpha \beta }^{\gamma },~\alpha ,\beta ,\gamma
\in \overline{1,p}$ will be called the \emph{structure functions of the
generalized Lie algebroid }$\left( \left( F,\nu ,N\right) ,\left[ ,\right]
_{F,h},\left( \rho ,\eta \right) \right) .$

\begin{itemize}
\item We assume the following diagrams:%
\begin{equation*}
\begin{array}[b]{ccccc}
F & ^{\underrightarrow{~\ \ \ \rho ~\ \ }} & TM & ^{\underrightarrow{~\ \ \
Th~\ \ }} & TN \\
~\downarrow \nu &  & ~\ \ \ \downarrow \tau _{M} &  & ~\ \ \ \downarrow \tau
_{N} \\
N & ^{\underrightarrow{~\ \ \ \eta ~\ \ }} & M & ^{\underrightarrow{~\ \ \
h~\ \ }} & N \\
&  &  &  &  \\
\left( \chi ^{\tilde{\imath}},z^{\alpha }\right) &  & \left(
x^{i},y^{i}\right) &  & \left( \chi ^{\tilde{\imath}},z^{\tilde{\imath}%
}\right)%
\end{array}%
\end{equation*}

where $i,\tilde{\imath}\in \overline{1,m}$ and $\alpha \in \overline{1,p}.$

If%
\begin{equation*}
\left( \chi ^{\tilde{\imath}},z^{\alpha }\right) \longrightarrow \left( \chi
^{\tilde{\imath}\prime }\left( \chi ^{\tilde{\imath}}\right) ,z^{\alpha
\prime }\left( \chi ^{\tilde{\imath}},z^{\alpha }\right) \right) ,
\end{equation*}%
\begin{equation*}
\left( x^{i},y^{i}\right) \longrightarrow \left( x^{i%
{\acute{}}%
}\left( x^{i}\right) ,y^{i%
{\acute{}}%
}\left( x^{i},y^{i}\right) \right)
\end{equation*}%
and
\begin{equation*}
\left( \chi ^{\tilde{\imath}},z^{\tilde{\imath}}\right) \longrightarrow
\left( \chi ^{\tilde{\imath}\prime }\left( \chi ^{\tilde{\imath}}\right) ,z^{%
\tilde{\imath}\prime }\left( \chi ^{\tilde{\imath}},z^{\tilde{\imath}%
}\right) \right) ,
\end{equation*}%
then
\begin{equation*}
\begin{array}[b]{c}
z^{\alpha
{\acute{}}%
}=\Lambda _{\alpha }^{\alpha
{\acute{}}%
}z^{\alpha }%
\end{array}%
,
\end{equation*}%
\begin{equation*}
\begin{array}[b]{c}
y^{i%
{\acute{}}%
}=\frac{\partial x^{i%
{\acute{}}%
}}{\partial x^{i}}y^{i}%
\end{array}%
\end{equation*}%
and
\begin{equation*}
\begin{array}{c}
z^{\tilde{\imath}\prime }=\frac{\partial \chi ^{\tilde{\imath}\prime }}{%
\partial \chi ^{\tilde{\imath}}}z^{\tilde{\imath}}.%
\end{array}%
\end{equation*}

\item We assume that $\left( \theta ,\mu \right) \overset{put}{=}\left(
Th\circ \rho ,h\circ \eta \right) $. If $z^{\alpha }t_{\alpha }\in \Gamma
\left( F,\nu ,N\right) $ is arbitrary, then
\begin{equation*}
\begin{array}[t]{l}
\begin{array}{c}
\Gamma \left( Th\circ \rho ,h\circ \eta \right) \left( z^{\alpha }t_{\alpha
}\right) f\left( h\circ \eta \left( \varkappa \right) \right) =\vspace*{1mm}
\\
=\left( \theta _{\alpha }^{\tilde{\imath}}z^{\alpha }\frac{\partial f}{%
\partial \varkappa ^{\tilde{\imath}}}\right) \left( h\circ \eta \left(
\varkappa \right) \right) =\left( \left( \rho _{\alpha }^{i}\circ h\right)
\left( z^{\alpha }\circ h\right) \frac{\partial f\circ h}{\partial x^{i}}%
\right) \left( \eta \left( \varkappa \right) \right) ,%
\end{array}%
\end{array}%
\leqno(2.1)
\end{equation*}%
for any $f\in \mathcal{F}\left( N\right) $ and $\varkappa \in N.$
\end{itemize}

The coefficients $\rho _{\alpha }^{i}$ respectively $\theta _{\alpha }^{%
\tilde{\imath}}$ change to $\rho _{\alpha
{\acute{}}%
}^{i%
{\acute{}}%
}$ respectively $\theta _{\alpha
{\acute{}}%
}^{\tilde{\imath}%
{\acute{}}%
}$ according to the rule:
\begin{equation*}
\begin{array}{c}
\rho _{\alpha
{\acute{}}%
}^{i%
{\acute{}}%
}=\Lambda _{\alpha
{\acute{}}%
}^{\alpha }\rho _{\alpha }^{i}\displaystyle\frac{\partial x^{i%
{\acute{}}%
}}{\partial x^{i}},%
\end{array}%
\leqno(2.2)
\end{equation*}%
respectively%
\begin{equation*}
\begin{array}{c}
\theta _{\alpha
{\acute{}}%
}^{\tilde{\imath}%
{\acute{}}%
}=\Lambda _{\alpha
{\acute{}}%
}^{\alpha }\theta _{\alpha }^{\tilde{\imath}}\displaystyle\frac{\partial
\varkappa ^{\tilde{\imath}%
{\acute{}}%
}}{\partial \varkappa ^{\tilde{\imath}}},%
\end{array}%
\leqno(2.3)
\end{equation*}%
where
\begin{equation*}
\left\Vert \Lambda _{\alpha
{\acute{}}%
}^{\alpha }\right\Vert =\left\Vert \Lambda _{\alpha }^{\alpha
{\acute{}}%
}\right\Vert ^{-1}.
\end{equation*}

\emph{Remark 2.1 } The following equalities hold good:%
\begin{equation*}
\begin{array}{c}
\displaystyle\rho _{\alpha }^{i}\circ h\frac{\partial f\circ h}{\partial
x^{i}}=\left( \theta _{\alpha }^{\tilde{\imath}}\frac{\partial f}{\partial
\varkappa ^{\tilde{\imath}}}\right) \circ h,\forall f\in \mathcal{F}\left(
N\right) .%
\end{array}%
\leqno(2.4)
\end{equation*}%
\emph{and }%
\begin{equation*}
\begin{array}{c}
\displaystyle\left( L_{\alpha \beta }^{\gamma }\circ h\right) \left( \rho
_{\gamma }^{k}\circ h\right) =\left( \rho _{\alpha }^{i}\circ h\right) \frac{%
\partial \left( \rho _{\beta }^{k}\circ h\right) }{\partial x^{i}}-\left(
\rho _{\beta }^{j}\circ h\right) \frac{\partial \left( \rho _{\alpha
}^{k}\circ h\right) }{\partial x^{j}}.%
\end{array}%
\leqno(2.5)
\end{equation*}

\section{Legendre transformation}

Let $\left( E,\pi ,M\right) $ be a vector bundle. We take $\left(
x^{i},y^{a}\right) $ as canonical local coordinates on $\left( E,\pi
,M\right) ,$ where $i\in \overline{1,m}$ and $a\in \overline{1,r}.$ Consider
\begin{equation*}
\left( x^{i},y^{a}\right) \longrightarrow \left( x^{i%
{\acute{}}%
}\left( x^{i}\right) ,y^{a%
{\acute{}}%
}\left( x^{i},y^{a}\right) \right)
\end{equation*}%
a change of coordinates on $\left( E,\pi ,M\right) $. Then the coordinates $%
y^{a}$ change to $y^{a%
{\acute{}}%
}$ according to the rule:
\begin{equation*}
\begin{array}{c}
y^{a%
{\acute{}}%
}=M_{a}^{a%
{\acute{}}%
}y^{a}.%
\end{array}%
\leqno(3.1)
\end{equation*}

Let $\left( \partial _{i},\overset{\cdot }{\partial }_{a}\right) $ be the
natural base of the Lie $\mathcal{F}\left( E\right) $-algebra $\left( \Gamma
\left( TE,\tau _{E},E\right) ,+,\cdot ,\left[ ,\right] _{TE}\right) .$

Let $\left( \overset{\ast }{E},\overset{\ast }{\pi },M\right) $ be the dual
vector bundle of $\left( E,\pi ,M\right) $. We take $\left(
x^{i},p_{a}\right) $ as canonical local coordinates on $\left( \overset{\ast
}{E},\overset{\ast }{\pi },M\right) ,$ where $i\in \overline{1,m}$ and $a\in
\overline{1,r}.$ Consider
\begin{equation*}
\left( x^{i},p_{a}\right) \longrightarrow \left( x^{i%
{\acute{}}%
}\left( x^{i}\right) ,p_{a%
{\acute{}}%
}\left( x^{i},p_{a}\right) \right)
\end{equation*}%
a change of coordinates on $\left( \overset{\ast }{E},\overset{\ast }{\pi }%
,M\right) $. Then the coordinates $p_{a}$ change to $p_{a%
{\acute{}}%
}$ according to the rule:
\begin{equation*}
\begin{array}{c}
p_{a%
{\acute{}}%
}=M_{a%
{\acute{}}%
}^{a}p_{a}.%
\end{array}%
\leqno(3.2)
\end{equation*}

If $\left( U,s_{U}\right) $ and $\left( U,\overset{\ast }{s}_{U}\right) $
are vector local $\left( m+r\right) $-charts then
\begin{equation*}
\begin{array}{c}
M_{a%
{\acute{}}%
}^{a}\left( x\right) \cdot M_{b}^{a%
{\acute{}}%
}\left( x\right) =\delta _{b}^{a},~\forall x\in U.%
\end{array}%
\end{equation*}

Let $\left( \overset{\ast }{\partial }_{i},\overset{\cdot }{\partial }%
^{a}\right) $ be the natural base of the Lie $\mathcal{F}\left( \overset{%
\ast }{E}\right) $-algebra $\left( \Gamma \left( T\overset{\ast }{E},\tau _{%
\overset{\ast }{E}},\overset{\ast }{E}\right) ,+,\cdot ,\left[ ,\right] _{T%
\overset{\ast }{E}}\right) .$

Let $L$ be a Lagrangian on the total space of the vector bundle $\left(
E,\pi ,M\right) .$ (see $\left[ 4\right] $) If $\left( U,s_{U}\right) $ is a
vector local $\left( m+r\right) $-chart for $\left( E,\pi ,M\right) $, then
we obtain the following real functions defined on $\pi ^{-1}\left( U\right) $%
:%
\begin{equation*}
\begin{array}{cc}
L_{i}\overset{put}{=}\frac{\partial L}{\partial x^{i}} & L_{ib}\overset{put}{%
=}\frac{\partial ^{2}L}{\partial x^{i}\partial y^{b}} \\
L_{a}\overset{put}{=}\frac{\partial L}{\partial y^{a}} & L_{ab}\overset{put}{%
=}\frac{\partial ^{2}L}{\partial y^{a}\partial y^{b}}%
\end{array}%
.\leqno(3.3)
\end{equation*}

We build the fiber bundle morphism%
\begin{equation*}
\begin{array}{rcl}
E & ^{\underrightarrow{~\ \ \varphi _{L}~\ \ }} & \overset{\ast }{E} \\
\pi \downarrow &  & \downarrow \overset{\ast }{\pi } \\
M & ^{\underrightarrow{~\ \ Id_{M}~\ \ }} & M%
\end{array}%
,
\end{equation*}%
where $\ \varphi _{L}$ is locally defined
\begin{equation*}
\begin{array}{ccc}
\pi ^{-1}\left( U\right) & ^{\underrightarrow{~\ \ \ \varphi _{L}~\ \ }} &
\overset{\ast }{\pi }^{-1}\left( U\right) \\
u_{x} & \longmapsto & L_{a}\left( u_{x}\right) s^{a}\left( x\right)%
\end{array}%
,\leqno(3.4)
\end{equation*}%
for any vector local $\left( m+r\right) $-chart $\left( U,s_{U}\right) $ of $%
\left( E,\pi ,M\right) $ and for any vector local $\left( m+r\right) $-chart
$\left( U,\overset{\ast }{s}_{U}\right) $ of $\left( \overset{\ast }{E},%
\overset{\ast }{\pi },M\right) .$

We obtain the Hamiltonian $H,$ locally defined by
\begin{equation*}
\begin{array}{ccc}
\overset{\ast }{\pi }^{-1}\left( U\right) & ^{\underrightarrow{~\ \ H~\ \ }}
& \mathbb{R} \\
\overset{\ast }{u}_{x}=p_{a}s^{a} & \longmapsto & p_{a}y^{a}-L\left(
u_{x}\right)%
\end{array}%
,\leqno(3.5)
\end{equation*}%
for any vector local $\left( m+r\right) $-chart $\left( U,\overset{\ast }{s}%
_{U}\right) $ of $\left( \overset{\ast }{E},\overset{\ast }{\pi },M\right) ,$
where $\left( y^{a},~a\in \overline{1,r}\right) $ are the components of the
solutions of the differentiable equations%
\begin{equation*}
\begin{array}{c}
p_{b}=L_{b}\left( u_{x}\right) ,~u_{x}\in \pi ^{-1}\left( U\right) .%
\end{array}%
\end{equation*}

The Hamiltonian given by $\left( 3.5\right) $ will be called the \emph{%
Legendre transformation of the Lagrangian} $L.$

If $\left( U,\overset{\ast }{s}_{U}\right) $ is a vector local $\left(
m+r\right) $-chart for $\left( \overset{\ast }{E},\overset{\ast }{\pi }%
,M\right) $, then we obtain the following real functions defined on $\overset%
{\ast }{\pi }^{-1}\left( U\right) $:%
\begin{equation*}
\begin{array}{cc}
H_{i}=\frac{\partial H}{\partial x^{i}} & H_{i}^{b}=\frac{\partial ^{2}H}{%
\partial x^{i}\partial p_{b}} \\
H^{a}=\frac{\partial H}{\partial p_{a}} & H^{ab}=\frac{\partial ^{2}H}{%
\partial p_{a}\partial p_{b}}%
\end{array}%
.\leqno(3.3)^{\prime }
\end{equation*}

Using this Hamiltonian, we build the fiber bundle morphism%
\begin{equation*}
\begin{array}{rcl}
\overset{\ast }{E} & ^{\underrightarrow{~\ \ \ \varphi _{H}~\ \ }} & E \\
\overset{\ast }{\pi }\downarrow &  & \downarrow \pi \\
M & ^{\underrightarrow{~\ \ Id_{M}~\ \ }} & M%
\end{array}%
,
\end{equation*}%
where $\ \varphi _{H}$ is locally defined%
\begin{equation*}
\begin{array}{ccc}
\overset{\ast }{\pi }^{-1}\left( U\right) & ^{\underrightarrow{~\ \ \
\varphi _{H}~\ \ }} & \pi ^{-1}\left( U\right) \\
\overset{\ast }{u}_{x} & \longmapsto & H^{a}\left( \overset{\ast }{u}%
_{x}\right) s_{a}\left( x\right)%
\end{array}%
,\leqno(3.4)^{\prime }
\end{equation*}%
for any vector local $\left( m+r\right) $-chart $\left( U,s_{U}\right) $ of $%
\left( E,\pi ,M\right) $ and for any vector local $\left( m+r\right) $-chart
$\left( U,\overset{\ast }{s}_{U}\right) $ of $\left( \overset{\ast }{E},%
\overset{\ast }{\pi },M\right) .$

We obtain that the Lagrangian $L,$ is locally defined by
\begin{equation*}
\begin{array}{ccc}
\pi ^{-1}\left( U\right) & ^{\underrightarrow{~\ \ L~\ \ }} & \mathbb{R} \\
u_{x}=y^{a}s_{a} & \longmapsto & y^{a}p_{a}-H\left( \overset{\ast }{u}%
_{x}\right)%
\end{array}%
,\leqno(3.5)^{\prime }
\end{equation*}%
for any vector local $\left( m+r\right) $-chart $\left( U,s_{U}\right) $ of $%
\left( E,\pi ,M\right) ,$ where $\left( p_{a},~a\in \overline{1,r}\right) $
are the components of the solutions of the differentiable equations%
\begin{equation*}
\begin{array}{c}
y^{a}=H^{a}\left( \overset{\ast }{u}_{x}\right) ,~\overset{\ast }{u}_{x}\in
\overset{\ast }{\pi }^{-1}\left( U\right) .%
\end{array}%
\end{equation*}

We will say that $L$ is the \emph{Legendre transformation of the Hamiltonian}
$H.$

\emph{Remark 3.1 }For any vector local $\left( m+r\right) $-chart $\left(
U,s_{U}\right) $\ of $\left( E,\pi ,M\right) $\ and for any vector local $%
\left( m+r\right) $-chart $\left( U,\overset{\ast }{s}_{U}\right) $\ of $%
\left( \overset{\ast }{E},\overset{\ast }{\pi },M\right) $\ we obtain:
\begin{equation*}
\begin{array}{c}
\varphi _{H}\circ \varphi _{L}=Id_{\pi ^{-1}\left( U\right) }%
\end{array}%
\leqno(3.6)
\end{equation*}%
and
\begin{equation*}
\begin{array}{c}
\varphi _{L}\circ \varphi _{H}=Id_{\overset{\ast }{\pi }^{-1}\left( U\right)
}.%
\end{array}%
\leqno(3.7)
\end{equation*}

Therefore, locally, $\varphi _{L}$\ is diffeomorphism and $\varphi
_{L}^{-1}=\varphi _{H}.$

\section{Duality between Lie algebroids structures}

Using the diagram:%
\begin{equation*}
\begin{array}{rcl}
E &  & \left( F,\left[ ,\right] _{F,h},\left( \rho ,\eta \right) \right)  \\
\pi \downarrow  &  & ~\downarrow \nu  \\
M & ^{\underrightarrow{~\ \ \ \ h~\ \ \ \ }} & ~\ N%
\end{array}%
,\leqno(4.1)
\end{equation*}%
where $\left( \left( F,\nu ,N\right) ,\left[ ,\right] _{F,h},\left( \rho
,\eta \right) \right) $ is a generalized Lie algebroid, we build the Lie
algebroid generalized tangent bundle (see $\left[ 1,4\right] $)
\begin{equation*}
\begin{array}{c}
\left( \left( \left( \rho ,\eta \right) TE,\left( \rho ,\eta \right) \tau
_{E},E\right) ,\left[ ,\right] _{\left( \rho ,\eta \right) TE},\left( \tilde{%
\rho},Id_{E}\right) \right)
\end{array}%
.\leqno(4.2)
\end{equation*}

The natural $\left( \rho ,\eta \right) $-base of sections is denoted
\begin{equation*}
\begin{array}{c}
\left( \tilde{\partial}_{\alpha },\overset{\cdot }{\tilde{\partial}}%
_{a}\right) .%
\end{array}%
\leqno(4.3)
\end{equation*}

The Lie bracket $\left[ ,\right] _{\left( \rho ,\eta \right) TE}$ is defined
by%
\begin{equation*}
\begin{array}{l}
\left[ \left( Z_{1}^{\alpha }\tilde{\partial}_{\alpha }+Y_{1}^{a}\overset{%
\cdot }{\tilde{\partial}}_{a}\right) ,\left( Z_{2}^{\beta }\tilde{\partial}%
_{\beta }+Y_{2}^{b}\overset{\cdot }{\tilde{\partial}}_{b}\right) \right]
_{\left( \rho ,\eta \right) TE}=\vspace*{1mm} \\
=\left[ Z_{1}^{\alpha }T_{a},Z_{2}^{\beta }T_{\beta }\right] _{\pi ^{\ast
}\left( h^{\ast }F\right) }\oplus \left[ \left( \rho _{\alpha }^{i}\circ
h\circ \pi \right) Z_{1}^{\alpha }\partial _{i}+Y_{1}^{a}\dot{\partial}%
_{a},\right. \\
\hfill \left. \left( \rho _{\beta }^{j}\circ h\circ \pi \right) Z_{2}^{\beta
}\partial _{j}+Y_{2}^{b}\dot{\partial}_{b}\right] _{TE},%
\end{array}%
\leqno(4.4)
\end{equation*}%
for any sections $\left( Z_{1}^{\alpha }\tilde{\partial}_{\alpha }+Y_{1}^{a}%
\overset{\cdot }{\tilde{\partial}}_{a}\right) $ and $\left( Z_{2}^{\beta }%
\tilde{\partial}_{\beta }+Y_{2}^{b}\overset{\cdot }{\tilde{\partial}}%
_{b}\right) .$

The anchor map $\left( \tilde{\rho},Id_{E}\right) $\ is a $\mathbf{B}^{%
\mathbf{v}}$-morphism of $\left( \left( \rho ,\eta \right) TE,\left( \rho
,\eta \right) \tau _{E},E\right) $\ source and $\left( TE,\tau _{E},E\right)
$\ target, where
\begin{equation*}
\begin{array}{rcl}
\left( \rho ,\eta \right) TE & \!\!^{\underrightarrow{\tilde{\rho}}}\!\!\! &
\!\!TE \\
\left( Z^{\alpha }\tilde{\partial}_{\alpha }+Y^{a}\overset{\cdot }{\tilde{%
\partial}}_{a}\right) \!(u_{x})\!\!\! & \!\!\longmapsto \!\!\! &
\!\!\!\left( \!Z^{\alpha }\!\!\left( \rho _{\alpha }^{i}{\circ }h{\circ }\pi
\right) \partial _{i}{+}Y^{a}\dot{\partial}_{a}\right) \!(u_{x})\!\!.%
\end{array}%
\bigskip \leqno(4.5)
\end{equation*}

Using the diagram:
\begin{equation*}
\begin{array}{rcl}
\overset{\ast }{E} &  & \left( F,\left[ ,\right] _{F,h},\left( \rho ,\eta
\right) \right)  \\
\overset{\ast }{\pi }\downarrow  &  & ~\downarrow \nu  \\
M & ^{\underrightarrow{~\ \ \ \ h~\ \ \ \ }} & ~N%
\end{array}%
,\leqno(4.1)^{\prime }
\end{equation*}%
where $\left( \left( F,\nu ,N\right) ,\left[ ,\right] _{F,h},\left( \rho
,\eta \right) \right) $ is a generalized Lie algebroid, we build the Lie
algebroid generalized tangent bundle (see $\left[ 1,5\right] $)%
\begin{equation*}
\begin{array}{c}
\left( \left( \left( \rho ,\eta \right) T\overset{\ast }{E},\left( \rho
,\eta \right) \tau _{\overset{\ast }{E}},\overset{\ast }{E}\right) ,\left[ ,%
\right] _{\left( \rho ,\eta \right) T\overset{\ast }{E}},\left( \overset{%
\ast }{\tilde{\rho}},Id_{\overset{\ast }{E}}\right) \right) .%
\end{array}%
\leqno(4.2)^{\prime }
\end{equation*}

The natural $\left( \rho ,\eta \right) $-base of sections is denoted%
\begin{equation*}
\begin{array}{c}
\left( \overset{\ast }{\tilde{\partial}}_{\alpha },\overset{\cdot }{\tilde{%
\partial}}^{a}\right) .%
\end{array}%
\leqno(4.3)^{\prime }
\end{equation*}

The Lie bracket $\left[ ,\right] _{\left( \rho ,\eta \right) T\overset{\ast }%
{E}}$ is defined by%
\begin{equation*}
\begin{array}{l}
\left[ \left( Z_{1}^{\alpha }\overset{\ast }{\tilde{\partial}}_{\alpha
}+Y_{a}^{1}\overset{\cdot }{\tilde{\partial}}^{a}\right) ,\left(
Z_{2}^{\beta }\overset{\ast }{\tilde{\partial}}_{\beta }+Y_{b}^{2}\overset{%
\cdot }{\tilde{\partial}}^{b}\right) \right] _{\left( \rho ,\eta \right) T%
\overset{\ast }{E}}=\vspace*{1mm} \\
=\left[ Z_{1}^{\alpha }T_{a},Z_{2}^{\beta }T_{\beta }\right] _{\overset{\ast
}{\pi }^{^{\ast }}\left( h^{\ast }F\right) }\oplus \left[ \left( \rho
_{\alpha }^{i}\circ h\circ \overset{\ast }{\pi }\right) Z_{1}^{\alpha }%
\overset{\ast }{\partial }_{i}+Y_{a}^{1}\dot{\partial}^{a},\right. \\
\hfill \left. \left( \rho _{\beta }^{j}\circ h\circ \overset{\ast }{\pi }%
\right) Z_{2}^{\beta }\overset{\ast }{\partial }_{j}+Y_{b}^{2}\dot{\partial}%
^{b}\right] _{T\overset{\ast }{E}},%
\end{array}%
\leqno(4.4)^{\prime }
\end{equation*}%
for any sections $\left( Z_{1}^{\alpha }\overset{\ast }{\tilde{\partial}}%
_{\alpha }+Y_{a}^{1}\overset{\cdot }{\tilde{\partial}}^{a}\right) $ and $%
\left( Z_{2}^{\beta }\overset{\ast }{\tilde{\partial}}_{\beta }+Y_{b}^{2}%
\overset{\cdot }{\tilde{\partial}}^{b}\right) .$

The anchor map $\left( \overset{\ast }{\tilde{\rho}},Id_{\overset{\ast }{E}%
}\right) $\ is a $\mathbf{B}^{\mathbf{v}}$-morphism of $\left( \left( \rho
,\eta \right) T\overset{\ast }{E},\left( \rho ,\eta \right) \tau _{\overset{%
\ast }{E}},\overset{\ast }{E}\right) $\ source and $\left( T\overset{\ast }{E%
},\tau _{\overset{\ast }{E}},\overset{\ast }{E}\right) $\ target, where
\begin{equation*}
\begin{array}{rcl}
\left( \rho ,\eta \right) T\overset{\ast }{E}\!\!\! & \!\!^{\underrightarrow{%
\overset{\ast }{\tilde{\rho}}}}\!\!\! & \!\!T\overset{\ast }{E} \\
\!\left( Z^{\alpha }\overset{\ast }{\tilde{\partial}}_{\alpha }+Y_{a}\overset%
{\cdot }{\tilde{\partial}}^{a}\right) \!(\overset{\ast }{u_{x}})\!\!\!\! &
\!\!\longmapsto \!\!\! & \!\!\left( Z^{\alpha }\left( \rho _{\alpha }^{i}{%
\circ }h{\circ }\overset{\ast }{\pi }\!\right) \!\!\!\overset{\ast }{%
\partial }_{i}{+}Y_{a}\dot{\partial}^{a}\right) \!(\overset{\ast }{u}%
_{x})\!\!.%
\end{array}%
\bigskip \leqno(4.5)^{\prime }
\end{equation*}

Using the $\mathbf{B}$-morphism $\left( \varphi _{L},Id_{M}\right) $, we
build the $\mathbf{B}^{\mathbf{v}}$-morphism $\left( \left( \rho ,\eta
\right) T\varphi _{L},\varphi _{L}\right) $ given by the diagram%
\begin{equation*}
\begin{array}{rcl}
\left( \rho ,\eta \right) TE & ^{\underrightarrow{~\ \left( \rho ,\eta
\right) T\varphi _{L}~\ }} & \left( \rho ,\eta \right) T\overset{\ast }{E}
\\
\left( \rho ,\eta \right) \tau _{E}\downarrow &  & \downarrow \left( \rho
,\eta \right) \tau _{\overset{\ast }{E}} \\
E & ^{\underrightarrow{~\ \ \ \ \varphi _{L~\ \ \ \ }}} & \overset{\ast }{E}%
\end{array}%
,\leqno(4.6)
\end{equation*}%
such that
\begin{equation*}
\begin{array}{cl}
\Gamma \left( \left( \rho ,\eta \right) T\varphi _{L},\varphi _{L}\right)
\left( Z^{\alpha }\tilde{\partial}_{\alpha }\right) & =\left( Z^{\alpha
}\circ \varphi _{H}\right) \overset{\ast }{\tilde{\partial}}_{\alpha }+\left[
\left( \rho _{\alpha }^{i}{\circ }h{\circ }\pi \right) Z^{\alpha }L_{ib}%
\right] \circ \varphi _{H}\overset{\cdot }{\tilde{\partial}}^{b}, \\
\Gamma \left( \left( \rho ,\eta \right) T\varphi _{L},\varphi _{L}\right)
\left( Y^{a}\overset{\cdot }{\tilde{\partial}}_{a}\right) & =\left(
Y^{a}L_{ab}\right) \circ \varphi _{H}\overset{\cdot }{\tilde{\partial}}^{b},%
\end{array}%
\leqno(4.7)
\end{equation*}%
for any $Z^{\alpha }\tilde{\partial}_{\alpha }+Y^{a}\overset{\cdot }{\tilde{%
\partial}}_{a}\in \Gamma \left( \left( \rho ,\eta \right) TE,\left( \rho
,\eta \right) \tau _{E},E\right) ,$ where $H$ is the Legendre transformation
of the Lagrangian $L.$

The $\mathbf{B}^{\mathbf{v}}$-morphism $\left( \left( \rho ,\eta \right)
T\varphi _{L},\varphi _{L}\right) $ will be called the $\left( \rho ,\eta
\right) $\emph{-tangent application of the Legendre bundle morphism
associated to the Lagrangian }$L$.

Using the $\mathbf{B}$-morphism $\left( \varphi _{H},Id_{M}\right) $, we
build the $\mathbf{B}^{\mathbf{v}}$-morphism $\left( \left( \rho ,\eta
\right) T\varphi _{H},\varphi _{H}\right) $ given by the diagram
\begin{equation*}
\begin{array}{rcl}
\left( \rho ,\eta \right) T\overset{\ast }{E} & ^{\underrightarrow{~\ \left(
\rho ,\eta \right) T\varphi _{H}~\ }} & \left( \rho ,\eta \right) TE \\
\left( \rho ,\eta \right) \tau _{\overset{\ast }{E}}\downarrow &  &
\downarrow \left( \rho ,\eta \right) \tau _{E} \\
E^{\ast } & ^{\underrightarrow{~\ \ \ \ \varphi _{H~\ \ \ \ }}} & E,%
\end{array}%
\leqno(4.6)^{\prime }
\end{equation*}%
such that
\begin{equation*}
\begin{array}{cl}
\Gamma \left( \left( \rho ,\eta \right) T\varphi _{L},\varphi _{L}\right)
\left( Z^{\alpha }\overset{\ast }{\tilde{\partial}}_{\alpha }\right) &
=\left( Z^{\alpha }\circ \varphi _{L}\right) \tilde{\partial}_{\alpha }+%
\left[ \left( \rho _{\alpha }^{i}{\circ }h{\circ }\overset{\ast }{\pi }%
\right) Z^{\alpha }H_{i}^{b}\right] \circ \varphi _{L}\overset{\cdot }{%
\tilde{\partial}}_{b}, \\
\Gamma \left( \left( \rho ,\eta \right) T\varphi _{L},\varphi _{L}\right)
\left( Y_{a}\overset{\cdot }{\tilde{\partial}}^{a}\right) & =\left(
Y_{a}H^{ab}\right) \circ \varphi _{L}\overset{\cdot }{\tilde{\partial}}_{b},%
\end{array}%
\leqno(4.7)^{\prime }
\end{equation*}%
for any $Z^{\alpha }\overset{\ast }{\tilde{\partial}}_{\alpha }+Y_{a}\overset%
{\cdot }{\tilde{\partial}}^{a}\in \Gamma \left( \left( \rho ,\eta \right) T%
\overset{\ast }{E},\left( \rho ,\eta \right) \tau _{\overset{\ast }{E}},%
\overset{\ast }{E}\right) .$

The $\mathbf{B}^{\mathbf{v}}$-morphism $\left( \left( \rho ,\eta \right)
T\varphi _{H},\varphi _{H}\right) $ will be called the $\left( \rho ,\eta
\right) $\emph{-tangent application of the Legendre bundle morphism
associated to the Hamiltonian }$H$.

\textbf{Theorem 4.1 }\emph{If the} $\mathbf{B}^{\mathbf{v}}$\emph{-morphism }%
$\left( \left( \rho ,\eta \right) T\varphi _{L},\varphi _{L}\right) $ \emph{%
is morphism of Lie algebroids, then we obtain:}
\begin{equation*}
\begin{array}{c}
\left( L_{\alpha \beta }^{\gamma }\circ h\circ \pi \right) \circ \varphi
_{H}=L_{\alpha \beta }^{\gamma }\circ h\circ \overset{\ast }{\pi },%
\end{array}%
\leqno(4.8)
\end{equation*}%
\begin{equation*}
\begin{array}{cl}
^{\left[ \left( L_{\alpha \beta }^{\gamma }\rho _{\gamma }^{k}\right) \circ
h\circ \pi \cdot L_{kb}\right] \circ \varphi _{H}} & ^{=\rho _{\alpha }^{i}{%
\circ }h{\circ }\overset{\ast }{\pi }\cdot \frac{\partial }{\partial x^{i}}%
\left[ \left( \rho _{\beta }^{j}{\circ }h{\circ }\pi \cdot L_{jb}\right)
\circ \varphi _{H}\right] } \\
& ^{-\rho _{\beta }^{j}{\circ }h{\circ }\overset{\ast }{\pi }\cdot \frac{%
\partial }{\partial x^{j}}\left[ \left( \rho _{\alpha }^{i}{\circ }h{\circ }%
\pi \cdot L_{ib}\right) \circ \varphi _{H}\right] } \\
& ^{+\left( \rho _{\alpha }^{i}{\circ }h{\circ }\pi \cdot L_{ia}\right)
\circ \varphi _{H}\cdot \frac{\partial }{\partial p_{a}}\left[ \left( \rho
_{\beta }^{j}{\circ }h{\circ }\pi \cdot L_{jb}\right) \circ \varphi _{H}%
\right] } \\
& ^{-\left( \rho _{\beta }^{j}{\circ }h{\circ }\pi \cdot L_{ja}\right) \circ
\varphi _{H}\cdot \frac{\partial }{\partial p_{a}}\left[ \left( \rho
_{\alpha }^{i}{\circ }h{\circ }\pi \cdot L_{ib}\right) \circ \varphi _{H}%
\right] ,}%
\end{array}%
\leqno(4.9)
\end{equation*}%
\begin{equation*}
\begin{array}{cl}
0 & =\rho _{\alpha }^{i}{\circ }h{\circ }\overset{\ast }{\pi }\cdot \frac{%
\partial }{\partial x^{i}}\left( L_{ba}\circ \varphi _{H}\right) \\
& +\left( \rho _{\alpha }^{i}{\circ }h{\circ }\pi \cdot L_{bc}\right) \circ
\varphi _{H}\frac{\partial }{\partial p_{c}}\left( L_{ba}\circ \varphi
_{H}\right) \\
& -L_{bc}\circ \varphi _{H}\cdot \frac{\partial }{\partial p_{c}}\left[
\left( \rho _{\alpha }^{i}{\circ }h{\circ }\pi \cdot L_{ia}\right) \circ
\varphi _{H}\right]%
\end{array}%
\leqno(4.10)
\end{equation*}%
\emph{and}%
\begin{equation*}
\begin{array}{cl}
0 & =L_{ac}\circ \varphi _{H}\cdot \frac{\partial }{\partial p_{c}}\left(
L_{bd}\circ \varphi _{H}\right) \\
& -L_{bc}\circ \varphi _{H}\cdot \frac{\partial }{\partial p_{c}}\left(
L_{ad}\circ \varphi _{H}\right) .%
\end{array}%
\leqno\left( 4.11\right)
\end{equation*}

\bigskip\noindent\textit{Proof.} Developing the following equalities
\begin{equation*}
\begin{array}{c}
\Gamma \left( \left( \rho ,\eta \right) T\varphi _{L},\varphi _{L}\right)
\left[ \tilde{\partial}_{\alpha },\tilde{\partial}_{\beta }\right] _{\left(
\rho ,\eta \right) TE} \\
=\left[ \Gamma \left( \left( \rho ,\eta \right) T\varphi _{L},\varphi
_{L}\right) \tilde{\partial}_{\alpha },\Gamma \left( \left( \rho ,\eta
\right) T\varphi _{L},\varphi _{L}\right) \tilde{\partial}_{\beta }\right]
_{\left( \rho ,\eta \right) T\overset{\ast }{E}},%
\end{array}%
\end{equation*}%
\begin{equation*}
\begin{array}{c}
\Gamma \left( \left( \rho ,\eta \right) T\varphi _{L},\varphi _{L}\right)
\left[ \tilde{\partial}_{\alpha },\overset{\cdot }{\tilde{\partial}}_{b}%
\right] _{\left( \rho ,\eta \right) TE} \\
=\left[ \Gamma \left( \left( \rho ,\eta \right) T\varphi _{L},\varphi
_{L}\right) \tilde{\partial}_{\alpha },\Gamma \left( \left( \rho ,\eta
\right) T\varphi _{L},\varphi _{L}\right) \overset{\cdot }{\tilde{\partial}}%
_{b}\right] _{\left( \rho ,\eta \right) T\overset{\ast }{E}}%
\end{array}%
\end{equation*}%
and
\begin{equation*}
\begin{array}{c}
\Gamma \left( \left( \rho ,\eta \right) T\varphi _{L},\varphi _{L}\right)
\left[ \overset{\cdot }{\tilde{\partial}}_{a},\overset{\cdot }{\tilde{%
\partial}}_{b}\right] _{\left( \rho ,\eta \right) TE} \\
=\left[ \Gamma \left( \left( \rho ,\eta \right) T\varphi _{L},\varphi
_{L}\right) \overset{\cdot }{\tilde{\partial}}_{a},\Gamma \left( \left( \rho
,\eta \right) T\varphi _{L},\varphi _{L}\right) \overset{\cdot }{\tilde{%
\partial}}_{b}\right] _{\left( \rho ,\eta \right) T\overset{\ast }{E}}%
\end{array}%
\end{equation*}%
it results the conclusion of the theorem. \hfill \emph{q.e.d.}

\textbf{Corollary 4.1 }\emph{In particular case of Lie algebroids, }$\left(
\eta ,h\right) =\left( Id_{M},Id_{M}\right) $\emph{, we obtain:}%
\begin{equation*}
\begin{array}{c}
\left( L_{\alpha \beta }^{\gamma }\circ \pi \right) \circ \varphi
_{H}=L_{\alpha \beta }^{\gamma }\circ \overset{\ast }{\pi },%
\end{array}%
\leqno(4.8)^{\prime }
\end{equation*}%
\begin{equation*}
\begin{array}{cl}
^{\left[ \left( L_{\alpha \beta }^{\gamma }\rho _{\gamma }^{k}\right) \circ
\pi \cdot L_{kb}\right] \circ \varphi _{H}} & ^{=\rho _{\alpha }^{i}{\circ }%
\overset{\ast }{\pi }\cdot \frac{\partial }{\partial x^{i}}\left[ \left(
\rho _{\beta }^{j}{\circ }\pi \cdot L_{jb}\right) \circ \varphi _{H}\right] }
\\
& ^{-\rho _{\beta }^{j}{\circ }\overset{\ast }{\pi }\cdot \frac{\partial }{%
\partial x^{j}}\left[ \left( \rho _{\alpha }^{i}{\circ }\pi \cdot
L_{ib}\right) \circ \varphi _{H}\right] } \\
& ^{+\left( \rho _{\alpha }^{i}{\circ }\pi \cdot L_{ia}\right) \circ \varphi
_{H}\cdot \frac{\partial }{\partial p_{a}}\left[ \left( \rho _{\beta }^{j}{%
\circ }\pi \cdot L_{jb}\right) \circ \varphi _{H}\right] } \\
& ^{-\left( \rho _{\beta }^{j}{\circ }\pi \cdot L_{ja}\right) \circ \varphi
_{H}\cdot \frac{\partial }{\partial p_{a}}\left[ \left( \rho _{\alpha }^{i}{%
\circ }\pi \cdot L_{ib}\right) \circ \varphi _{H}\right] ,}%
\end{array}%
\leqno(4.9)^{\prime }
\end{equation*}%
\begin{equation*}
\begin{array}{cl}
0 & =\rho _{\alpha }^{i}{\circ }\overset{\ast }{\pi }\cdot \frac{\partial }{%
\partial x^{i}}\left( L_{ba}\circ \varphi _{H}\right) \\
& +\left( \rho _{\alpha }^{i}{\circ }\pi \cdot L_{bc}\right) \circ \varphi
_{H}\frac{\partial }{\partial p_{c}}\left( L_{ba}\circ \varphi _{H}\right)
\\
& -L_{bc}\circ \varphi _{H}\cdot \frac{\partial }{\partial p_{c}}\left[
\left( \rho _{\alpha }^{i}{\circ }\pi \cdot L_{ia}\right) \circ \varphi _{H}%
\right]%
\end{array}%
\leqno(4.10)^{\prime }
\end{equation*}%
\emph{and}%
\begin{equation*}
\begin{array}{cl}
0 & =L_{ac}\circ \varphi _{H}\cdot \frac{\partial }{\partial p_{c}}\left(
L_{bd}\circ \varphi _{H}\right) \\
& -L_{bc}\circ \varphi _{H}\cdot \frac{\partial }{\partial p_{c}}\left(
L_{ad}\circ \varphi _{H}\right) .%
\end{array}%
\leqno\left( 4.11\right) ^{\prime }
\end{equation*}

\emph{In the classical case, }$\left( \rho ,\eta ,h\right) =\left(
Id_{TM},Id_{M},Id_{M}\right) $\emph{, we obtain:}
\begin{equation*}
\begin{array}{cl}
0 & =\frac{\partial }{\partial x^{i}}\left( \frac{\partial ^{2}L}{\partial
x^{j}\partial y^{k}}\circ \varphi _{H}\right) -\frac{\partial }{\partial
x^{j}}\left( \frac{\partial ^{2}L}{\partial x^{i}\partial y^{k}}\circ
\varphi _{H}\right) \\
& +\frac{\partial ^{2}L}{\partial x^{i}\partial y^{h}}\circ \varphi
_{H}\cdot \frac{\partial }{\partial p_{h}}\left( \frac{\partial ^{2}L}{%
\partial x^{j}\partial y^{k}}\circ \varphi _{H}\right) -\frac{\partial ^{2}L%
}{\partial x^{j}\partial y^{h}}\circ \varphi _{H}\cdot \frac{\partial }{%
\partial p_{h}}\left( \frac{\partial ^{2}L}{\partial x^{i}\partial y^{k}}%
\circ \varphi _{H}\right)%
\end{array}%
\leqno(4.9)^{\prime \prime }
\end{equation*}%
\begin{equation*}
\begin{array}{cl}
0 & =\frac{\partial }{\partial x^{i}}\left( \frac{\partial ^{2}L}{\partial
y^{j}\partial y^{k}}\circ \varphi _{H}\right) +\frac{\partial ^{2}L}{%
\partial x^{i}\partial y^{h}}\circ \varphi _{H}\cdot \frac{\partial }{%
\partial p_{h}}\left( \frac{\partial ^{2}L}{\partial y^{j}\partial y^{k}}%
\circ \varphi _{H}\right) \\
& -\frac{\partial ^{2}L}{\partial x^{j}\partial y^{h}}\circ \varphi
_{H}\cdot \frac{\partial }{\partial p_{h}}\left( \frac{\partial ^{2}L}{%
\partial x^{i}\partial y^{k}}\circ \varphi _{H}\right)%
\end{array}%
\leqno(4.10)^{\prime \prime }
\end{equation*}%
\emph{and}%
\begin{equation*}
\begin{array}{cl}
0 & =\frac{\partial ^{2}L}{\partial y^{i}\partial y^{h}}\circ \varphi
_{H}\cdot \frac{\partial }{\partial p_{h}}\left( \frac{\partial ^{2}L}{%
\partial y^{j}\partial y^{k}}\circ \varphi _{H}\right) \\
& -\frac{\partial ^{2}L}{\partial y^{j}\partial y^{h}}\circ \varphi
_{H}\cdot \frac{\partial }{\partial p_{h}}\left( \frac{\partial ^{2}L}{%
\partial y^{i}\partial y^{k}}\circ \varphi _{H}\right) .%
\end{array}%
\leqno(4.11)^{\prime \prime }
\end{equation*}

\textbf{Theorem 4.2 }\emph{Dual, if the }$\mathbf{B}^{\mathbf{v}}$\emph{%
-morphism} $\left( \left( \rho ,\eta \right) T\varphi _{H},\varphi
_{H}\right) $ \emph{is morphism of Lie algebroids, then we obtain:}
\begin{equation*}
\begin{array}{c}
\left( L_{\alpha \beta }^{\gamma }\circ h\circ \overset{\ast }{\pi }\right)
\circ \varphi _{L}=L_{\alpha \beta }^{\gamma }\circ h\circ \pi ,%
\end{array}%
\leqno(4.12)
\end{equation*}%
\begin{equation*}
\begin{array}{cl}
^{\left[ \left( L_{\alpha \beta }^{\gamma }\rho _{\gamma }^{k}\right) \circ
h\circ \overset{\ast }{\pi }\cdot H_{k}^{b}\right] \circ \varphi _{L}} &
^{=\rho _{\alpha }^{i}{\circ }h{\circ }\pi \cdot \frac{\partial }{\partial
x^{i}}\left[ \left( \rho _{\beta }^{j}{\circ }h{\circ }\overset{\ast }{\pi }%
\cdot H_{j}^{b}\right) \circ \varphi _{L}\right] } \\
& ^{-\rho _{\beta }^{j}{\circ }h{\circ }\pi \cdot \frac{\partial }{\partial
x^{j}}\left[ \left( \rho _{\alpha }^{i}{\circ }h{\circ }\overset{\ast }{\pi }%
\cdot H_{i}^{b}\right) \circ \varphi _{L}\right] } \\
& ^{+\left( \rho _{\alpha }^{i}{\circ }h{\circ }\overset{\ast }{\pi }\cdot
H_{i}^{c}\right) \circ \varphi _{L}\cdot \frac{\partial }{\partial y^{c}}%
\left[ \left( \rho _{\beta }^{j}{\circ }h{\circ }\overset{\ast }{\pi }\cdot
H_{j}^{b}\right) \circ \varphi _{L}\right] } \\
& ^{-\left( \rho _{\beta }^{j}{\circ }h{\circ }\overset{\ast }{\pi }\cdot
H_{j}^{c}\right) \circ \varphi _{L}\cdot \frac{\partial }{\partial y^{c}}%
\left[ \left( \rho _{\alpha }^{i}{\circ }h{\circ }\overset{\ast }{\pi }\cdot
H_{i}^{b}\right) \circ \varphi _{L}\right] ,}%
\end{array}%
\leqno(4.13)
\end{equation*}%
\begin{equation*}
\begin{array}{cl}
0 & =\rho _{\alpha }^{i}{\circ }h{\circ }\pi \cdot \frac{\partial }{\partial
x^{i}}\left( H^{ba}\circ \varphi _{L}\right) \\
& +\left( \rho _{\alpha }^{i}{\circ }h{\circ }\overset{\ast }{\pi }\cdot
H^{bc}\right) \circ \varphi _{L}\frac{\partial }{\partial y^{c}}\left(
H^{ba}\circ \varphi _{L}\right) \\
& -H^{bc}\circ \varphi _{L}\cdot \frac{\partial }{\partial y^{c}}\left[
\left( \rho _{\alpha }^{i}{\circ }h{\circ }\overset{\ast }{\pi }\cdot
H_{i}^{a}\right) \circ \varphi _{L}\right]%
\end{array}%
\leqno(4.14)
\end{equation*}%
\emph{and}%
\begin{equation*}
\begin{array}{cl}
0 & =H^{ac}\circ \varphi _{L}\cdot \frac{\partial }{\partial y^{c}}\left(
H^{bd}\circ \varphi _{L}\right) \\
& -H^{bc}\circ \varphi _{L}\cdot \frac{\partial }{\partial y^{c}}\left(
H^{ad}\circ \varphi _{L}\right) .%
\end{array}%
\leqno(4.15)
\end{equation*}

\bigskip \noindent \textit{Proof.} Developing the following equalities
\begin{equation*}
\begin{array}{c}
\Gamma \left( \left( \rho ,\eta \right) T\varphi _{H},\varphi _{H}\right)
\left[ \overset{\ast }{\tilde{\partial}}_{\alpha },\overset{\ast }{\tilde{%
\partial}}_{\beta }\right] _{\left( \rho ,\eta \right) T\overset{\ast }{E}}
\\
=\left[ \Gamma \left( \left( \rho ,\eta \right) T\varphi _{H},\varphi
_{H}\right) \overset{\ast }{\tilde{\partial}}_{\alpha },\Gamma \left( \left(
\rho ,\eta \right) T\varphi _{H},\varphi _{H}\right) \overset{\ast }{\tilde{%
\partial}}_{\beta }\right] _{\left( \rho ,\eta \right) TE},%
\end{array}%
\end{equation*}%
\begin{equation*}
\begin{array}{c}
\Gamma \left( \left( \rho ,\eta \right) T\varphi _{H},\varphi _{H}\right)
\left[ \overset{\ast }{\tilde{\partial}}_{\alpha },\overset{\cdot }{\tilde{%
\partial}}^{b}\right] _{\left( \rho ,\eta \right) T\overset{\ast }{E}} \\
=\left[ \Gamma \left( \left( \rho ,\eta \right) T\varphi _{H},\varphi
_{H}\right) \overset{\ast }{\tilde{\partial}}_{\alpha },\Gamma \left( \left(
\rho ,\eta \right) T\varphi _{H},\varphi _{H}\right) \overset{\cdot }{\tilde{%
\partial}}^{b}\right] _{\left( \rho ,\eta \right) TE}%
\end{array}%
\end{equation*}%
and
\begin{equation*}
\begin{array}{c}
\Gamma \left( \left( \rho ,\eta \right) T\varphi _{H},\varphi _{H}\right)
\left[ \overset{\cdot }{\tilde{\partial}}^{a},\overset{\cdot }{\tilde{%
\partial}}^{b}\right] _{\left( \rho ,\eta \right) T\overset{\ast }{E}} \\
=\left[ \Gamma \left( \left( \rho ,\eta \right) T\varphi _{H},\varphi
_{H}\right) \overset{\cdot }{\tilde{\partial}}^{a},\Gamma \left( \left( \rho
,\eta \right) T\varphi _{H},\varphi _{H}\right) \overset{\cdot }{\tilde{%
\partial}}^{b}\right] _{\left( \rho ,\eta \right) TE}%
\end{array}%
\end{equation*}%
it results the conclusion of the theorem. \hfill \emph{q.e.d.}\bigskip

\textbf{Corollary 4.2 }\emph{In the particular case of Lie algebroids, }$%
\left( \eta ,h\right) =\left( Id_{M},Id_{M}\right) $\emph{, we obtain:}%
\begin{equation*}
\begin{array}{c}
\left( L_{\alpha \beta }^{\gamma }\circ \overset{\ast }{\pi }\right) \circ
\varphi _{L}=L_{\alpha \beta }^{\gamma }\circ \pi ,%
\end{array}%
\leqno(4.12)^{\prime }
\end{equation*}%
\begin{equation*}
\begin{array}{cl}
^{\left[ \left( L_{\alpha \beta }^{\gamma }\rho _{\gamma }^{k}\right) \circ
\overset{\ast }{\pi }\cdot H_{k}^{b}\right] \circ \varphi _{L}} & ^{=\rho
_{\alpha }^{i}{\circ }\pi \cdot \frac{\partial }{\partial x^{i}}\left[
\left( \rho _{\beta }^{j}{\circ }\overset{\ast }{\pi }\cdot H_{j}^{b}\right)
\circ \varphi _{L}\right] } \\
& ^{-\rho _{\beta }^{j}{\circ }\pi \cdot \frac{\partial }{\partial x^{j}}%
\left[ \left( \rho _{\alpha }^{i}{\circ }\overset{\ast }{\pi }\cdot
H_{i}^{b}\right) \circ \varphi _{L}\right] } \\
& ^{+\left( \rho _{\alpha }^{i}{\circ }\overset{\ast }{\pi }\cdot
H_{i}^{c}\right) \circ \varphi _{L}\cdot \frac{\partial }{\partial y^{c}}%
\left[ \left( \rho _{\beta }^{j}{\circ }\overset{\ast }{\pi }\cdot
H_{j}^{b}\right) \circ \varphi _{L}\right] } \\
& ^{-\left( \rho _{\beta }^{j}{\circ }\overset{\ast }{\pi }\cdot
H_{j}^{c}\right) \circ \varphi _{L}\cdot \frac{\partial }{\partial y^{c}}%
\left[ \left( \rho _{\alpha }^{i}{\circ }\overset{\ast }{\pi }\cdot
H_{i}^{b}\right) \circ \varphi _{L}\right] ,}%
\end{array}%
\leqno(4.13)^{\prime }
\end{equation*}%
\begin{equation*}
\begin{array}{cl}
0 & =\rho _{\alpha }^{i}{\circ }\pi \cdot \frac{\partial }{\partial x^{i}}%
\left( H^{ba}\circ \varphi _{L}\right) \\
& +\left( \rho _{\alpha }^{i}{\circ }\overset{\ast }{\pi }\cdot
H^{bc}\right) \circ \varphi _{L}\frac{\partial }{\partial y^{c}}\left(
H^{ba}\circ \varphi _{L}\right) \\
& -H^{bc}\circ \varphi _{L}\cdot \frac{\partial }{\partial y^{c}}\left[
\left( \rho _{\alpha }^{i}{\circ }\overset{\ast }{\pi }\cdot
H_{i}^{a}\right) \circ \varphi _{L}\right]%
\end{array}%
\leqno(4.14)^{\prime }
\end{equation*}%
\emph{and}%
\begin{equation*}
\begin{array}{cl}
0 & =H^{ac}\circ \varphi _{L}\cdot \frac{\partial }{\partial y^{c}}\left(
H^{bd}\circ \varphi _{L}\right) \\
& -H^{bc}\circ \varphi _{L}\cdot \frac{\partial }{\partial y^{c}}\left(
H^{ad}\circ \varphi _{L}\right) .%
\end{array}%
\leqno(4.15)^{\prime }
\end{equation*}

\emph{In the classical case, }$\left( \rho ,\eta ,h\right) =\left(
Id_{TM},Id_{M},Id_{M}\right) $\emph{, we obtain:}%
\begin{equation*}
(4.13)^{\prime \prime }%
\begin{array}{cl}
0 & =\frac{\partial }{\partial x^{i}}\left( \frac{\partial ^{2}H}{\partial
x^{k}\partial p_{j}}\circ \varphi _{L}\right) -\frac{\partial }{\partial
x^{k}}\left( \frac{\partial ^{2}H}{\partial x^{i}\partial p_{j}}\circ
\varphi _{L}\right) \\
& +\frac{\partial ^{2}H}{\partial x^{i}\partial p_{h}}\circ \varphi
_{L}\cdot \frac{\partial }{\partial y^{h}}\left( \frac{\partial ^{2}H}{%
\partial x^{k}\partial p_{j}}\circ \varphi _{L}\right) -\frac{\partial ^{2}H%
}{\partial x^{k}\partial p_{h}}\circ \varphi _{L}\cdot \frac{\partial }{%
\partial y^{h}}\left( \left( \frac{\partial ^{2}H}{\partial x^{i}\partial
p_{j}}\circ \varphi _{L}\right) \circ \varphi _{L}\right)%
\end{array}%
\end{equation*}%
\begin{equation*}
\begin{array}{cl}
0 & =\frac{\partial }{\partial x^{k}}\left( \frac{\partial ^{2}H}{\partial
p_{i}\partial p_{j}}\circ \varphi _{L}\right) +\frac{\partial ^{2}H}{%
\partial p_{i}\partial p_{h}}\circ \varphi _{L}\cdot \frac{\partial }{%
\partial y^{h}}\left( \frac{\partial ^{2}H}{\partial x_{k}\partial p_{j}}%
\circ \varphi _{L}\right) \\
& -\frac{\partial ^{2}H}{\partial p_{j}\partial p_{h}}\circ \varphi
_{L}\cdot \frac{\partial }{\partial y^{h}}\left( \frac{\partial ^{2}H}{%
\partial x^{k}\partial p_{i}}\circ \varphi _{L}\right)%
\end{array}%
\leqno(4.14)^{\prime \prime }
\end{equation*}%
\emph{and}%
\begin{equation*}
\begin{array}{cl}
0 & =\frac{\partial ^{2}H}{\partial p_{i}\partial p_{k}}\circ \varphi
_{L}\cdot \frac{\partial }{\partial y^{k}}\left( \frac{\partial ^{2}H}{%
\partial p_{j}\partial p_{h}}\circ \varphi _{L}\right) \\
& -\frac{\partial ^{2}H}{\partial p_{j}\partial p_{k}}\circ \varphi
_{L}\cdot \frac{\partial }{\partial y^{k}}\left( \frac{\partial ^{2}H}{%
\partial p_{i}\partial p_{h}}\circ \varphi _{L}\right) .%
\end{array}%
\leqno(4.15)^{\prime \prime }
\end{equation*}

\textbf{Definition 4.1 }If $\left( \left( \rho ,\eta \right) T\varphi
_{L},\varphi _{L}\right) $ and $\left( \left( \rho ,\eta \right) T\varphi
_{H},\varphi _{H}\right) $ are Lie algebroids morphisms, then we will say
that $\left( E,\pi ,M\right) $\emph{\ and }$\left( \overset{\ast }{E},%
\overset{\ast }{\pi },M\right) $\emph{\ are Legendre }$\left( \rho ,\eta
,h\right) $\emph{-equivalent} and we will write%
\begin{equation*}
\begin{array}{c}
\left( E,\pi ,M\right) \overset{\mathcal{L}}{\widetilde{_{\left( \rho ,\eta
,h\right) }}}\left( \overset{\ast }{E},\overset{\ast }{\pi },M\right) .%
\end{array}%
\end{equation*}

\section{Duality between adapted $\left( \protect\rho ,\protect\eta \right) $%
-basis}

If $\left( \rho ,\eta \right) \Gamma $ is a $\left( \rho ,\eta \right) $%
-connection for the vector bundle $\left( E,\pi ,M\right) ,$ then
\begin{equation*}
\begin{array}{c}
\left( \tilde{\partial}_{\alpha }-\left( \rho ,\eta \right) \Gamma _{\alpha
}^{a}\overset{\cdot }{\tilde{\partial}}_{a},\overset{\cdot }{\tilde{\partial}%
}_{a}\right) =\left( \tilde{\delta}_{\alpha },\overset{\cdot }{\tilde{%
\partial}}_{a}\right) .%
\end{array}%
\leqno(5.1)
\end{equation*}%
is the adapted $\left( \rho ,\eta \right) $-base of $\left( \Gamma \left(
\left( \rho ,\eta \right) TE,\left( \rho ,\eta \right) \tau _{E},E\right)
,+,\cdot \right) $. (see $\left[ 1,4\right] $)

If $\left( \rho ,\eta \right) \Gamma $ is a $\left( \rho ,\eta \right) $%
-connection for the vector bundle $\left( \overset{\ast }{E},\overset{\ast }{%
\pi },M\right) ,$ then
\begin{equation*}
\begin{array}{c}
\left( \overset{\ast }{\tilde{\partial}}_{\alpha }+\left( \rho ,\eta \right)
\Gamma _{b\alpha }\overset{\cdot }{\tilde{\partial}}^{b},\overset{\cdot }{%
\tilde{\partial}}^{a}\right) =\left( \overset{\ast }{\tilde{\delta}}_{\alpha
},\overset{\cdot }{\tilde{\partial}}^{a}\right) .%
\end{array}%
\leqno(5.1)^{\prime }
\end{equation*}%
is the adapted $\left( \rho ,\eta \right) $-base of $\left( \Gamma \left(
\left( \rho ,\eta \right) T\overset{\ast }{E},\left( \rho ,\eta \right) \tau
_{\overset{\ast }{E}},\overset{\ast }{E}\right) ,+,\cdot \right) $. (see $%
\left[ 1,5\right] $)

\textbf{Definition 5.1 }If%
\begin{equation*}
\begin{array}{c}
\left( E,\pi ,M\right) \overset{\mathcal{L}}{\widetilde{_{\left( \rho ,\eta
,h\right) }}}\left( \overset{\ast }{E},\overset{\ast }{\pi },M\right) .%
\end{array}%
\end{equation*}%
and
\begin{equation*}
\begin{array}{c}
\Gamma \left( \left( \rho ,\eta \right) T\varphi _{L},\varphi _{L}\right)
\left( \tilde{\delta}_{\alpha }\right) =\overset{\ast }{\tilde{\delta}}%
_{\alpha },%
\end{array}%
\end{equation*}%
\begin{equation*}
\begin{array}{c}
\Gamma \left( \left( \rho ,\eta \right) T\varphi _{H},\varphi _{H}\right)
\left( \overset{\ast }{\tilde{\delta}}_{\alpha }\right) =\tilde{\delta}%
_{\alpha },%
\end{array}%
\end{equation*}%
then we will say that $\left( E,\pi ,M\right) $\emph{\ and }$\left( \overset{%
\ast }{E},\overset{\ast }{\pi },M\right) $\emph{\ are horizontal Legendre }$%
\left( \rho ,\eta ,h\right) $\emph{-equivalent }and we will write%
\begin{equation*}
\begin{array}{c}
\left( E,\pi ,M\right) \overset{\mathcal{HL}}{\widetilde{_{\left( \rho ,\eta
,h\right) }}}\left( \overset{\ast }{E},\overset{\ast }{\pi },M\right) .%
\end{array}%
\end{equation*}

\textbf{Theorem 5.1 }\emph{If}
\begin{equation*}
\begin{array}{c}
\left( E,\pi ,M\right) \overset{\mathcal{HL}}{\widetilde{_{\left( \rho ,\eta
,h\right) }}}\left( \overset{\ast }{E},\overset{\ast }{\pi },M\right) ,%
\end{array}%
\end{equation*}%
\emph{then we obtain:}%
\begin{equation*}
\begin{array}{c}
\left( \rho ,\eta \right) \Gamma _{b\alpha }=\left[ \left( \rho _{\alpha
}^{i}{\circ }h{\circ }\pi \right) \cdot L_{ib}-\left( \rho ,\eta \right)
\Gamma _{\alpha }^{a}\cdot L_{ab}\right] \circ \varphi _{H}%
\end{array}%
\leqno(5.2)
\end{equation*}%
\emph{and}%
\begin{equation*}
\begin{array}{c}
-\left( \rho ,\eta \right) \Gamma _{\alpha }^{a}=\left[ \left( \rho _{\alpha
}^{i}{\circ }h{\circ }\overset{\ast }{\pi }\right) \cdot H_{i}^{a}+\left(
\rho ,\eta \right) \Gamma _{b\alpha }\cdot H^{ba}\right] \circ \varphi _{L}.%
\end{array}%
\leqno(5.3)
\end{equation*}

\textbf{Corollary 5.1 }\emph{In the particular case of Lie algebroids,} $%
\left( \eta ,h\right) =\left( Id_{M},Id_{M}\right) $\emph{, we obtain}%
\begin{equation*}
\begin{array}{c}
\rho \Gamma _{b\alpha }=\left[ \left( \rho _{\alpha }^{i}{\circ }\pi \right)
\cdot L_{ib}-\rho \Gamma _{\alpha }^{a}\cdot L_{ab}\right] \circ \varphi _{H}%
\end{array}%
\leqno(5.2)^{\prime }
\end{equation*}%
\emph{and}%
\begin{equation*}
\begin{array}{c}
-\rho \Gamma _{\alpha }^{a}=\left[ \left( \rho _{\alpha }^{i}{\circ }\overset%
{\ast }{\pi }\right) \cdot H_{i}^{a}+\rho \Gamma _{b\alpha }\cdot H^{ba}%
\right] \circ \varphi _{L}.%
\end{array}%
\leqno(5.3)^{\prime }
\end{equation*}

\emph{In the classical case, }$\left( \rho ,\eta ,h\right) =\left(
Id_{TM},Id_{M},Id_{M}\right) $\emph{, we obtain the equality} \emph{implies
the equality}%
\begin{equation*}
\begin{array}{c}
\Gamma _{jk}=\left[ \frac{\partial ^{2}L}{\partial x^{i}\partial y^{j}}%
-\Gamma _{k}^{i}\frac{\partial ^{2}L}{\partial y^{i}\partial y^{j}}\right]
\circ \varphi _{H}%
\end{array}%
\leqno(5.2)^{\prime \prime }
\end{equation*}%
\emph{and}%
\begin{equation*}
\begin{array}{c}
-\Gamma _{k}^{i}=\left[ \frac{\partial ^{2}H}{\partial x^{k}\partial p_{i}}%
+\Gamma _{jk}\frac{\partial ^{2}H}{\partial p_{j}\partial p_{i}}\right]
\circ \varphi _{L}.%
\end{array}%
\leqno(5.3)^{\prime \prime }
\end{equation*}

If the Lagrangian $L$ is regular, then we will define the real local
functions $\tilde{L}^{ab}$ such that
\begin{equation*}
\begin{array}{c}
\left\Vert \tilde{L}^{ab}\left( u_{x}\right) \right\Vert =\left\Vert
L_{ab}\left( u_{x}\right) \right\Vert ^{-1},~\forall u_{x}\in \pi
^{-1}\left( U\right) .%
\end{array}%
\end{equation*}

If the Hamiltonian $H$ is regular, then we will define the real local
functions $\tilde{H}_{ab}$ such that
\begin{equation*}
\begin{array}{c}
\left\Vert \tilde{H}_{ab}\left( \overset{\ast }{u}_{x}\right) \right\Vert
=\left\Vert H^{ab}\left( \overset{\ast }{u}_{x}\right) \right\Vert
^{-1},~\forall \overset{\ast }{u}_{x}\in \overset{\ast }{\pi }^{-1}\left(
U\right) .%
\end{array}%
\end{equation*}

\emph{Remark 5.1}\textbf{\ }If the Lagrangian $L$ is regular and
\begin{equation*}
\begin{array}{c}
\left( E,\pi ,M\right) \overset{\mathcal{HL}}{\widetilde{_{\left( \rho ,\eta
,h\right) }}}\left( \overset{\ast }{E},\overset{\ast }{\pi },M\right)%
\end{array}%
\end{equation*}%
then, the Hamiltonian $H$ is regular,%
\begin{equation*}
\begin{array}[b]{c}
\tilde{H}_{ab}=L_{ab}\circ \varphi _{H}.%
\end{array}%
\leqno(5.4)
\end{equation*}%
and
\begin{equation*}
\begin{array}{c}
\left[ \left( \rho _{\alpha }^{i}{\circ }h{\circ }\pi \right) \cdot L_{ia}%
\right] \circ \varphi _{H}=-\left( \rho _{\alpha }^{i}{\circ }h{\circ }%
\overset{\ast }{\pi }\right) \cdot H_{i}^{b}\cdot \tilde{H}_{ba}.%
\end{array}%
\leqno(5.5)
\end{equation*}

It is known that the following equalities hold good\textbf{\ }
\begin{equation*}
\begin{array}{c}
\left[ \tilde{\delta}_{\alpha },\tilde{\delta}_{\beta }\right] _{\left( \rho
,\eta \right) TE}=\left( L_{\alpha \beta }^{\gamma }\circ h\circ \pi \right)
\tilde{\delta}_{\gamma }+\left( \rho ,\eta ,h\right) \mathbb{R}_{\,\ ~\alpha
\beta }^{a}\overset{\cdot }{\tilde{\partial}}_{a},%
\end{array}%
\leqno(5.6)
\end{equation*}%
and
\begin{equation*}
\begin{array}{c}
\left[ \overset{\ast }{\tilde{\delta}}_{\alpha },\overset{\ast }{\tilde{%
\delta}}_{\beta }\right] _{\left( \rho ,\eta \right) T\overset{\ast }{E}%
}=\left( L_{\alpha \beta }^{\gamma }\circ h\circ \overset{\ast }{\pi }%
\right) \overset{\ast }{\tilde{\delta}}_{\gamma }+\left( \rho ,\eta
,h\right) \mathbb{R}_{b\,\ \alpha \beta }\overset{\cdot }{\tilde{\partial}}%
^{b},%
\end{array}%
\leqno(5.6)^{\prime }
\end{equation*}

\textbf{Theorem 5.2 }\emph{If}
\begin{equation*}
\begin{array}{c}
\left( E,\pi ,M\right) \overset{\mathcal{HL}}{\widetilde{_{\left( \rho ,\eta
,h\right) }}}\left( \overset{\ast }{E},\overset{\ast }{\pi },M\right) ,%
\end{array}%
\end{equation*}%
\emph{then, we obtain:}%
\begin{equation*}
\begin{array}{c}
\left( \rho ,\eta ,h\right) \mathbb{R}_{b~\alpha \beta }=\left[ \left( \rho
,\eta ,h\right) \mathbb{R}_{~~\ \alpha \beta }^{a}\cdot L_{ab}\right] \circ
\varphi _{H}%
\end{array}%
\leqno(5.7)
\end{equation*}%
\emph{and}
\begin{equation*}
\begin{array}{c}
\left( \rho ,\eta ,h\right) \mathbb{R}_{~\ \ \alpha \beta }^{a}=\left[
\left( \rho ,\eta ,h\right) \mathbb{R}_{b~\alpha \beta }\cdot H^{ba}\right]
\circ \varphi _{L}.%
\end{array}%
\leqno(5.8)
\end{equation*}

\textbf{Corollary 5.2 }\emph{In the particular case of Lie algebroids,} $%
\left( \eta ,h\right) =\left( Id_{M},Id_{M}\right) $\emph{, we obtain}%
\begin{equation*}
\begin{array}{c}
\rho \mathbb{R}_{b~\alpha \beta }=\left( \rho \mathbb{R}_{~~\ \alpha \beta
}^{a}L_{ab}\right) \circ \varphi _{H}%
\end{array}%
\leqno(5.7)^{\prime }
\end{equation*}%
\emph{and}
\begin{equation*}
\begin{array}{c}
\rho \mathbb{R}_{~\ \ \alpha \beta }^{a}=\left( \rho \mathbb{R}_{b~\alpha
\beta }H^{ba}\right) \circ \varphi _{L}.%
\end{array}%
\leqno(5.8)^{\prime }
\end{equation*}

\emph{In the classical case, }$\left( \rho ,\eta ,h\right) =\left(
Id_{TM},Id_{M},Id_{M}\right) $\emph{, we obtain }%
\begin{equation*}
\begin{array}{c}
\mathbb{R}_{j~hk}=\left( \mathbb{R}_{~~\ hk}^{i}\cdot \frac{\partial ^{2}L}{%
\partial y^{i}\partial y^{j}}\right) \circ \varphi _{H}%
\end{array}%
\leqno(5.7)^{\prime \prime }
\end{equation*}%
\emph{and}
\begin{equation*}
\begin{array}{c}
\mathbb{R}_{~\ \ hk}^{i}=\left( \mathbb{R}_{j~hk}\cdot \frac{\partial ^{2}H}{%
\partial p_{j}\partial p_{i}}\right) \circ \varphi _{L}.%
\end{array}%
\leqno(5.8)^{\prime \prime }
\end{equation*}

\textbf{Theorem 5.3 }\emph{If}
\begin{equation*}
\begin{array}{c}
\left( E,\pi ,M\right) \overset{\mathcal{HL}}{\widetilde{_{\left( \rho ,\eta
,h\right) }}}\left( \overset{\ast }{E},\overset{\ast }{\pi },M\right) ,%
\end{array}%
\end{equation*}%
\emph{then we obtain}%
\begin{equation*}
\begin{array}{cl}
\left( \frac{\partial \left( \rho ,\eta \right) \Gamma _{\alpha }^{a}}{%
\partial y^{b}}\cdot L_{ac}\right) \circ \varphi _{H} & =L_{ba}\circ \varphi
_{H}\cdot \frac{\partial \left( \rho ,\eta \right) \Gamma _{c\alpha }}{%
\partial p_{a}} \\
&
\begin{array}{l}
+\left( \rho _{\alpha }^{i}{\circ }h{\circ }\overset{\ast }{\pi }\right)
\cdot \frac{\partial }{\partial x^{i}}\left( L_{bc}\circ \varphi _{H}\right)
\\
+\left( \rho ,\eta \right) \Gamma _{a\alpha }\cdot \frac{\partial }{\partial
p_{a}}\left( L_{bc}\circ \varphi _{H}\right)%
\end{array}%
\end{array}%
\leqno(5.9)
\end{equation*}%
\emph{and}
\begin{equation*}
\begin{array}{cl}
-\left( \frac{\partial \left( \rho ,\eta \right) \Gamma _{b\alpha }}{%
\partial p_{a}}\cdot H^{bc}\right) \circ \varphi _{L} & =H^{ba}\circ \varphi
_{L}\cdot \frac{\partial \left( \rho ,\eta \right) \Gamma _{\alpha }^{c}}{%
\partial y^{a}} \\
&
\begin{array}{l}
+\left( \rho _{\alpha }^{i}{\circ }h{\circ }\pi \right) \cdot \frac{\partial
}{\partial x^{i}}\left( H^{bc}\circ \varphi _{L}\right) \\
+\left( \rho ,\eta \right) \Gamma _{\alpha }^{a}\cdot \frac{\partial }{%
\partial y^{a}}\left( H^{bc}\circ \varphi _{L}\right)%
\end{array}%
\end{array}%
\leqno(5.10)
\end{equation*}

\bigskip \noindent \emph{Proof. }Developing the following equalities%
\begin{equation*}
\begin{array}{c}
\Gamma \left( \left( \rho ,\eta \right) T\varphi _{L},\varphi _{L}\right)
\left( \left[ \tilde{\delta}_{\alpha },\overset{\cdot }{\tilde{\partial}}_{a}%
\right] _{\left( \rho ,\eta \right) TE}\right) \\
=\left[ \Gamma \left( \left( \rho ,\eta \right) T\varphi _{L},\varphi
_{L}\right) \tilde{\delta}_{\alpha },\Gamma \left( \left( \rho ,\eta \right)
T\varphi _{L},\varphi _{L}\right) \overset{\cdot }{\tilde{\partial}}_{a}%
\right] _{\left( \rho ,\eta \right) T\overset{\ast }{E}}%
\end{array}%
\end{equation*}%
and
\begin{equation*}
\begin{array}{c}
\Gamma \left( \left( \rho ,\eta \right) T\varphi _{H},\varphi _{H}\right)
\left( \left[ \overset{\ast }{\tilde{\delta}_{\alpha }},\overset{\cdot }{%
\tilde{\partial}}^{a}\right] _{\left( \rho ,\eta \right) T\overset{\ast }{E}%
}\right) \\
=\left[ \Gamma \left( \left( \rho ,\eta \right) T\varphi _{H},\varphi
_{H}\right) \overset{\ast }{\tilde{\delta}_{\alpha }},\Gamma \left( \left(
\rho ,\eta \right) T\varphi _{H},\varphi _{H}\right) \overset{\cdot }{\tilde{%
\partial}}_{a}\right] _{\left( \rho ,\eta \right) TE}%
\end{array}%
\end{equation*}%
it results the conclusion of the theorem.\hfill \emph{q.e.d.}

\textbf{Corollary 5.2 }\emph{In the particular case of Lie algebroids,} $%
\left( \eta ,h\right) =\left( Id_{M},Id_{M}\right) $\emph{, we obtain}%
\begin{equation*}
\begin{array}{cl}
\left( \frac{\partial \rho \Gamma _{\alpha }^{a}}{\partial y^{b}}\cdot
L_{ac}\right) \circ \varphi _{H} & =L_{ba}\circ \varphi _{H}\cdot \frac{%
\partial \rho \Gamma _{c\alpha }}{\partial p_{a}} \\
&
\begin{array}{l}
+\left( \rho _{\alpha }^{i}{\circ }\overset{\ast }{\pi }\right) \cdot \frac{%
\partial }{\partial x^{i}}\left( L_{bc}\circ \varphi _{H}\right) \\
+\rho \Gamma _{a\alpha }\cdot \frac{\partial }{\partial p_{a}}\left(
L_{bc}\circ \varphi _{H}\right)%
\end{array}%
\end{array}%
\leqno(5.9)^{\prime }
\end{equation*}%
\emph{and}
\begin{equation*}
\begin{array}{cl}
-\left( \frac{\partial \rho \Gamma _{b\alpha }}{\partial p_{a}}\cdot
H^{bc}\right) \circ \varphi _{L} & =H^{ba}\circ \varphi _{L}\cdot \frac{%
\partial \rho \Gamma _{\alpha }^{c}}{\partial y^{a}} \\
&
\begin{array}{l}
+\left( \rho _{\alpha }^{i}{\circ }\pi \right) \cdot \frac{\partial }{%
\partial x^{i}}\left( H^{bc}\circ \varphi _{L}\right) \\
+\rho \Gamma _{\alpha }^{a}\cdot \frac{\partial }{\partial y^{a}}\left(
H^{bc}\circ \varphi _{L}\right)%
\end{array}%
\end{array}%
\leqno(5.10)^{\prime }
\end{equation*}

\emph{In the classical case, }$\left( \rho ,\eta ,h\right) =\left(
Id_{TM},Id_{M},Id_{M}\right) $\emph{, we obtain}%
\begin{equation*}
\begin{array}{cl}
\left( \frac{\partial \Gamma _{k}^{i}}{\partial y^{j}}\cdot \frac{\partial
^{2}L}{\partial y^{i}\partial y^{h}}\right) \circ \varphi _{H} & =\frac{%
\partial ^{2}L}{\partial y^{j}\partial y^{i}}\circ \varphi _{H}\cdot \frac{%
\partial \rho \Gamma _{hk}}{\partial p_{i}} \\
&
\begin{array}{l}
+\frac{\partial }{\partial x^{k}}\left( \frac{\partial ^{2}L}{\partial
y^{j}\partial y^{h}}\circ \varphi _{H}\right) \\
+\Gamma _{ik}\cdot \frac{\partial }{\partial p_{i}}\left( \frac{\partial
^{2}L}{\partial y^{j}\partial y^{h}}\circ \varphi _{H}\right)%
\end{array}%
\end{array}%
\leqno(5.9)^{\prime \prime }
\end{equation*}%
\emph{and}
\begin{equation*}
\begin{array}{cl}
-\left( \frac{\partial \Gamma _{jk}}{\partial p_{i}}\cdot \frac{\partial
^{2}H}{\partial p_{j}\partial p_{h}}\right) \circ \varphi _{L} & =\frac{%
\partial ^{2}H}{\partial p_{i}\partial p_{e}}\circ \varphi _{L}\cdot \frac{%
\partial \rho \Gamma _{k}^{h}}{\partial y^{e}} \\
&
\begin{array}{l}
+\frac{\partial }{\partial x^{k}}\left( \frac{\partial ^{2}H}{\partial
p_{j}\partial p_{h}}\circ \varphi _{L}\right) \\
+\Gamma _{k}^{i}\cdot \frac{\partial }{\partial y^{i}}\left( \frac{\partial
^{2}H}{\partial p_{j}\partial p_{h}}\circ \varphi _{L}\right)%
\end{array}%
\end{array}%
\leqno(5.10)^{\prime \prime }
\end{equation*}

The dual natural $\left( \rho ,\eta \right) $-base of the natural $\left(
\rho ,\eta \right) $-base $\left( \tilde{\partial}_{\alpha },\overset{\cdot }%
{\tilde{\partial}}_{a}\right) $ is denoted $\left( d\tilde{z}^{\alpha },d%
\tilde{y}^{a}\right) $ and the dual adapted $\left( \rho ,\eta \right) $%
-base of the adapted $\left( \rho ,\eta \right) $-base $\left( \tilde{\delta}%
_{\alpha },\overset{\cdot }{\tilde{\partial}}_{a}\right) $ is denoted
\begin{equation*}
\begin{array}{c}
\left( d\tilde{z}^{\alpha },\delta \tilde{y}^{a}\right) \overset{put}{=}%
\left( d\tilde{z}^{\alpha },\left( \rho ,\eta \right) \Gamma _{\alpha
}^{a}\cdot d\tilde{z}^{\alpha }+d\tilde{y}^{a}\right) .%
\end{array}%
\leqno(5.11)
\end{equation*}

The dual natural $\left( \rho ,\eta \right) $-base of the natural $\left(
\rho ,\eta \right) $-base $\left( \overset{\ast }{\tilde{\partial}}_{\alpha
},\overset{\cdot }{\tilde{\partial}}^{a}\right) $ is denoted $\left( d\tilde{%
z}^{\alpha },d\tilde{p}_{a}\right) $ and the dual adapted $\left( \rho ,\eta
\right) $-base of the adapted $\left( \rho ,\eta \right) $-base $\left(
\overset{\ast }{\tilde{\delta}}_{\alpha },\overset{\cdot }{\tilde{\partial}}%
_{a}\right) $ is denoted
\begin{equation*}
\begin{array}{c}
\left( d\tilde{z}^{\alpha },\delta \tilde{p}_{a}\right) \overset{put}{=}%
\left( d\tilde{z}^{\alpha },-\left( \rho ,\eta \right) \Gamma _{a\alpha
}\cdot d\tilde{z}^{\alpha }+d\tilde{p}_{a}\right) .%
\end{array}%
\leqno(5.11)^{\prime }
\end{equation*}

Let
\begin{equation*}
\left( \Lambda \left( \left( \rho ,\eta \right) TE,\left( \rho ,\eta \right)
\tau _{E},E\right) ,+,\cdot ,\wedge \right)
\end{equation*}%
be exterior differential $\mathcal{F}\left( E\right) $-algebra of the
generalized tangent bundle $\left( \left( \rho ,\eta \right) TE,\left( \rho
,\eta \right) \tau _{E},E\right) $ and let
\begin{equation*}
\left( \Lambda \left( \left( \rho ,\eta \right) T\overset{\ast }{E},\left(
\rho ,\eta \right) \tau _{\overset{\ast }{E}},\overset{\ast }{E}\right)
,+,\cdot ,\wedge \right) .
\end{equation*}%
be exterior differential $\mathcal{F}\left( \overset{\ast }{E}\right) $%
-algebra of the generalized tangent bundle $\left( \left( \rho ,\eta \right)
T\overset{\ast }{E},\left( \rho ,\eta \right) \tau _{\overset{\ast }{E}},%
\overset{\ast }{E}\right) .$

Using the $\mathbf{B}^{\mathbf{v}}$-morphism $\left( 4.6\right) $ given by
the equalities $\left( 4.7\right) $, we obtain the application
\begin{equation*}
\begin{array}[b]{ccc}
\Lambda \left( \left( \rho ,\eta \right) T\overset{\ast }{E},\left( \rho
,\eta \right) \tau _{\overset{\ast }{E}},\overset{\ast }{E}\right) & ^{%
\underrightarrow{~\left( \left( \rho ,\eta \right) T\varphi _{L},\varphi
_{L}\right) ^{\ast }\ \ }} & \Lambda \left( \left( \rho ,\eta \right)
TE,\left( \rho ,\eta \right) \tau _{E},E\right) \\
\Lambda ^{q}\left( \left( \rho ,\eta \right) T\overset{\ast }{E},\left( \rho
,\eta \right) \tau _{\overset{\ast }{E}},\overset{\ast }{E}\right) \ni \omega
& \longmapsto & \left( \left( \rho ,\eta \right) T\varphi _{L},\varphi
_{L}\right) ^{\ast }\left( \omega \right)%
\end{array}%
\end{equation*}%
where
\begin{equation*}
\left( \left( \rho ,\eta \right) T\varphi _{L},\varphi _{L}\right) ^{\ast
}\left( \omega \right) \left( X_{1},...,X_{q}\right) =\omega \left( \Gamma
\left( \left( \rho ,\eta \right) T\varphi _{L},\varphi _{L}\right)
X_{1},...,\Gamma \left( \left( \rho ,\eta \right) T\varphi _{L},\varphi
_{L}\right) X_{q}\right) \circ \varphi _{L},
\end{equation*}%
for any $X_{1},...,X_{q}\in \Gamma \left( \left( \rho ,\eta \right)
TE,\left( \rho ,\eta \right) \tau _{E},E\right) .$

Using the $\mathbf{B}^{\mathbf{v}}$-morphism $\left( 4.6\right) ^{\prime }$
given by the equalities $\left( 4.7\right) ^{\prime }$, we obtain the
application
\begin{equation*}
\begin{array}[b]{ccc}
\Lambda \left( \left( \rho ,\eta \right) TE,\left( \rho ,\eta \right) \tau
_{E},E\right) & ^{\underrightarrow{~\left( \left( \rho ,\eta \right)
T\varphi _{H},\varphi _{H}\right) ^{\ast }\ \ }} & \Lambda \left( \left(
\rho ,\eta \right) T\overset{\ast }{E},\left( \rho ,\eta \right) \tau _{%
\overset{\ast }{E}},\overset{\ast }{E}\right) \\
\Lambda ^{q}\left( \left( \rho ,\eta \right) TE,\left( \rho ,\eta \right)
\tau _{E},E\right) \ni \omega & \longmapsto & \left( \left( \rho ,\eta
\right) T\varphi _{H},\varphi _{H}\right) ^{\ast }\left( \omega \right)%
\end{array}%
\end{equation*}%
where
\begin{equation*}
\left( \left( \rho ,\eta \right) T\varphi _{H},\varphi _{H}\right) ^{\ast
}\left( \omega \right) \left( X_{1},...,X_{q}\right) =\omega \left( \Gamma
\left( \left( \rho ,\eta \right) T\varphi _{H},\varphi _{H}\right)
X_{1},...,\Gamma \left( \left( \rho ,\eta \right) T\varphi _{H},\varphi
_{H}\right) X_{q}\right) \circ \varphi _{H},
\end{equation*}%
for any $X_{1},...,X_{q}\in \Gamma \left( \left( \rho ,\eta \right) T\overset%
{\ast }{E},\left( \rho ,\eta \right) \tau _{\overset{\ast }{E}},\overset{%
\ast }{E}\right) .$

\textbf{Theorem 5.4 }\emph{The equality }%
\begin{equation*}
\begin{array}[b]{c}
\Gamma \left( \left( \rho ,\eta \right) T\varphi _{L},\varphi _{L}\right)
\left( \tilde{\delta}_{\alpha }\right) =\overset{\ast }{\tilde{\delta}}%
_{\alpha }%
\end{array}%
\end{equation*}%
\emph{\ is equivalent with the equality:}%
\begin{equation*}
\begin{array}{c}
\left( \left( \rho ,\eta \right) T\varphi _{L},\varphi _{L}\right) ^{\ast
}\left( \delta \tilde{p}_{a}\right) =L_{ab}\cdot \delta \tilde{y}^{b}.%
\end{array}%
\leqno(5.12)
\end{equation*}

\emph{Dual, the equality }%
\begin{equation*}
\begin{array}{c}
\Gamma \left( \left( \rho ,\eta \right) T\varphi _{H},\varphi _{H}\right)
\left( \overset{\ast }{\tilde{\delta}}_{\alpha }\right) =\tilde{\delta}%
_{\alpha },%
\end{array}%
\end{equation*}%
\emph{\ is equivalent with the equality:}%
\begin{equation*}
\begin{array}{c}
\Gamma \left( \left( \rho ,\eta \right) T\varphi _{H},\varphi _{H}\right)
^{\ast }\left( \delta \tilde{y}^{a}\right) =H^{ab}\cdot \delta \tilde{p}_{b}%
\end{array}%
\leqno(5.12)^{\prime }
\end{equation*}

\section{Duality between distinguished linear $\left( \protect\rho ,\protect%
\eta \right) $-connections}

Let $\left( \rho ,\eta \right) \Gamma $ be a $\left( \rho ,\eta \right) $%
-connection for the vector bundle $\left( E,\pi ,M\right) $ and let
\begin{equation*}
\begin{array}{l}
\left( X,T\right) ^{\underrightarrow{\left( \rho ,\eta \right) D}\,}\vspace*{%
1mm}\left( \rho ,\eta \right) D_{X}T%
\end{array}%
\leqno(6.1)
\end{equation*}%
be a covariant $\left( \rho ,\eta \right) $-derivative for the tensor
algebra of generalized tangent bundle
\begin{equation*}
\left( \left( \rho ,\eta \right) TE,\left( \rho ,\eta \right) \tau
_{E},E\right)
\end{equation*}%
which \ preserves \ the \ horizontal and vertical \emph{IDS} by parallelism.

If $\left( U,s_{U}\right) $ is a vector local $\left( m+r\right) $-chart for
$\left( E,\pi ,M\right) ,$ then the real local functions
\begin{equation*}
\left( \left( \rho ,\eta \right) H_{\beta \gamma }^{\alpha },\left( \rho
,\eta \right) H_{b\gamma }^{a},\left( \rho ,\eta \right) V_{\beta c}^{\alpha
},\left( \rho ,\eta \right) V_{bc}^{a}\right)
\end{equation*}%
defined on $\pi ^{-1}\left( U\right) $ and determined by the following
equalities:
\begin{equation*}
\begin{array}{ll}
\left( \rho ,\eta \right) D_{\tilde{\delta}_{\gamma }}\tilde{\delta}_{\beta
}=\left( \rho ,\eta \right) H_{\beta \gamma }^{\alpha }\tilde{\delta}%
_{\alpha }, & \left( \rho ,\eta \right) D_{\tilde{\delta}_{\gamma }}\overset{%
\cdot }{\tilde{\partial}}_{b}=\left( \rho ,\eta \right) H_{b\gamma }^{a}%
\overset{\cdot }{\tilde{\partial}}_{a} \\
\left( \rho ,\eta \right) D_{\overset{\cdot }{\tilde{\partial}}_{c}}\tilde{%
\delta}_{\beta }=\left( \rho ,\eta \right) V_{\beta _{c}}^{\alpha }\tilde{%
\delta}_{\alpha }, & \left( \rho ,\eta \right) D_{\overset{\cdot }{\tilde{%
\partial}}_{c}}\overset{\cdot }{\tilde{\partial}}_{b}=\left( \rho ,\eta
\right) V_{bc}^{a}\overset{\cdot }{\tilde{\partial}}_{a}%
\end{array}%
\leqno(6.2)
\end{equation*}%
are the components of a distinguished linear $\left( \rho ,\eta \right) $%
-connection $\left( \left( \rho ,\eta \right) H,\left( \rho ,\eta \right)
V\right) .$

Let $\left( \rho ,\eta \right) \Gamma $ be a $\left( \rho ,\eta \right) $%
-connection for the vector bundle $\left( \overset{\ast }{E},\overset{\ast }{%
\pi },M\right) $ and let
\begin{equation*}
\begin{array}{l}
\left( X,T\right) ^{\underrightarrow{\left( \rho ,\eta \right) \overset{\ast
}{D}}\,}\vspace*{1mm}\left( \rho ,\eta \right) \overset{\ast }{D}_{X}T%
\end{array}%
\leqno(6.1)^{\prime }
\end{equation*}%
be a covariant $\left( \rho ,\eta \right) $-derivative for the tensor
algebra of generalized tangent bundle
\begin{equation*}
\left( \left( \rho ,\eta \right) T\overset{\ast }{E},\left( \rho ,\eta
\right) \tau _{\overset{\ast }{E}},\overset{\ast }{E}\right)
\end{equation*}%
which \ preserves \ the \ horizontal and vertical \emph{IDS} by parallelism.

If $\left( U,\overset{\ast }{s}_{U}\right) $ is a vector local $\left(
m+r\right) $-chart for $\left( \overset{\ast }{E},\overset{\ast }{\pi }%
,M\right) ,$ then the real local functions
\begin{equation*}
\left( \left( \rho ,\eta \right) \overset{\ast }{H}_{\beta \gamma }^{\alpha
},\left( \rho ,\eta \right) \overset{\ast }{H}_{b\gamma }^{a},\left( \rho
,\eta \right) \overset{\ast }{V}_{\beta }^{\alpha c},\left( \rho ,\eta
\right) \overset{\ast }{V}_{b}^{ac}\right)
\end{equation*}%
defined on $\overset{\ast }{\pi }^{-1}\left( U\right) $ and determined by
the following equalities:
\begin{equation*}
\begin{array}{ll}
\left( \rho ,\eta \right) \overset{\ast }{D}_{\overset{\ast }{\tilde{\delta}}%
_{\gamma }}\overset{\ast }{\tilde{\delta}}_{\beta }=\left( \rho ,\eta
\right) \overset{\ast }{H}_{\beta \gamma }^{\alpha }\overset{\ast }{\tilde{%
\delta}}_{\alpha }, & \left( \rho ,\eta \right) \overset{\ast }{D}_{\overset{%
\ast }{\tilde{\delta}}_{\gamma }}\overset{\cdot }{\tilde{\partial}}%
^{a}=\left( \rho ,\eta \right) \overset{\ast }{H}_{b\gamma }^{a}\overset{%
\cdot }{\tilde{\partial}}^{b} \\
\left( \rho ,\eta \right) \overset{\ast }{D}_{\overset{\cdot }{\tilde{%
\partial}}^{c}}\overset{\ast }{\tilde{\delta}}_{\beta }=\left( \rho ,\eta
\right) \overset{\ast }{V}_{\beta }^{\alpha c}\overset{\ast }{\tilde{\delta}}%
_{\alpha }, & \left( \rho ,\eta \right) \overset{\ast }{D}_{\overset{\cdot }{%
\tilde{\partial}}^{c}}\overset{\cdot }{\tilde{\partial}}^{b}=\left( \rho
,\eta \right) \overset{\ast }{V}_{a}^{bc}\overset{\cdot }{\tilde{\partial}}%
^{a}%
\end{array}%
\leqno(6.2)^{\prime }
\end{equation*}%
are the components of a distinguished linear $\left( \rho ,\eta \right) $%
-connection
\begin{equation*}
\left( \left( \rho ,\eta \right) \overset{\ast }{H},\left( \rho ,\eta
\right) \overset{\ast }{V}\right) .
\end{equation*}

\textbf{Theorem 6.1 }\emph{If}
\begin{equation*}
\begin{array}{c}
\left( E,\pi ,M\right) \overset{\mathcal{HL}}{\widetilde{_{\left( \rho ,\eta
,h\right) }}}\left( \overset{\ast }{E},\overset{\ast }{\pi },M\right)%
\end{array}%
\end{equation*}%
and
\begin{equation*}
\begin{array}{c}
\Gamma \left( \left( \rho ,\eta \right) T\varphi _{L},\varphi _{L}\right)
\left( \left( \rho ,\eta \right) D_{X}Y\right) =\left( \rho ,\eta \right)
\overset{\ast }{D}_{\Gamma \left( \left( \rho ,\eta \right) T\varphi
_{L},\varphi _{L}\right) X}\Gamma \left( \left( \rho ,\eta \right) T\varphi
_{L},\varphi _{L}\right) Y,%
\end{array}%
\end{equation*}%
for any $X,Y\in \Gamma \left( \left( \rho ,\eta \right) TE,\left( \rho ,\eta
\right) \tau _{E},E\right) $, then we obtain:%
\begin{equation*}
\begin{array}{cc}
\left( \rho ,\eta \right) H_{\beta \gamma }^{\alpha }\circ \varphi _{H} &
=\left( \rho ,\eta \right) \overset{\ast }{H}_{\beta \gamma }^{\alpha },%
\end{array}%
\leqno(6.3)
\end{equation*}%
\begin{equation*}
\begin{array}{rl}
\left( \left( \rho ,\eta \right) H_{b\gamma }^{a}\cdot L_{ac}\right) \circ
\varphi _{H} & =\left( \rho _{\gamma }^{k}{\circ }h{\circ }\overset{\ast }{%
\pi }\right) \cdot \frac{\partial }{\partial x^{k}}\left( L_{bc}\circ
\varphi _{H}\right) \\
& +\left( \rho ,\eta \right) \Gamma _{b\gamma }\cdot \frac{\partial }{%
\partial p_{b}}\left( L_{bc}\circ \varphi _{H}\right) \\
& -\left( \rho ,\eta \right) \overset{\ast }{H}_{b\gamma }^{a}\cdot \left(
L_{ac}\circ \varphi _{H}\right) ,%
\end{array}%
\leqno(6.4)
\end{equation*}%
\begin{equation*}
\begin{array}{cc}
\left( \rho ,\eta \right) V_{\beta d}^{\alpha }\circ \varphi _{H} & =\left(
\rho ,\eta \right) \overset{\ast }{V}_{\beta }^{\alpha c}\cdot \left(
L_{cd}\circ \varphi _{H}\right)%
\end{array}%
\leqno(6.5)
\end{equation*}%
and%
\begin{equation*}
\begin{array}{rl}
\left( \left( \rho ,\eta \right) V_{bc}^{a}\cdot L_{ad}\right) \circ \varphi
_{H} & =\left( L_{ce}\circ \varphi _{H}\right) \cdot \frac{\partial }{%
\partial p_{e}}\left( L_{bd}\circ \varphi _{H}\right) \\
& -\left( L_{ce}\circ \varphi _{H}\right) \cdot \left( \rho ,\eta \right)
\overset{\ast }{V}_{d}^{ef}\cdot \left( L_{bf}\circ \varphi _{H}\right) .%
\end{array}%
\leqno(6.6)
\end{equation*}

\textbf{Corollary 6.1 }\emph{In the particular case of Lie algebroids,} $%
\left( \eta ,h\right) =\left( Id_{M},Id_{M}\right) $\emph{, we obtain}%
\begin{equation*}
\begin{array}{cc}
\rho H_{\beta \gamma }^{\alpha }\circ \varphi _{H} & =\rho \overset{\ast }{H}%
_{\beta \gamma }^{\alpha },%
\end{array}%
\leqno(6.3)^{\prime }
\end{equation*}%
\begin{equation*}
\begin{array}{rl}
\left( \rho H_{b\gamma }^{a}\cdot L_{ac}\right) \circ \varphi _{H} & =\left(
\rho _{\gamma }^{k}{\circ }\overset{\ast }{\pi }\right) \cdot \frac{\partial
}{\partial x^{k}}\left( L_{bc}\circ \varphi _{H}\right) \\
& +\rho \Gamma _{b\gamma }\cdot \frac{\partial }{\partial p_{b}}\left(
L_{bc}\circ \varphi _{H}\right) \\
& -\rho \overset{\ast }{H}_{b\gamma }^{a}\cdot \left( L_{ac}\circ \varphi
_{H}\right) ,%
\end{array}%
\leqno(6.4)^{\prime }
\end{equation*}%
\begin{equation*}
\begin{array}{cc}
\rho V_{\beta d}^{\alpha }\circ \varphi _{H} & =\rho \overset{\ast }{V}%
_{\beta }^{\alpha c}\cdot \left( L_{cd}\circ \varphi _{H}\right)%
\end{array}%
\leqno(6.5)^{\prime }
\end{equation*}%
and%
\begin{equation*}
\begin{array}{rl}
\left( \rho V_{bc}^{a}\cdot L_{ad}\right) \circ \varphi _{H} & =\left(
L_{ce}\circ \varphi _{H}\right) \cdot \frac{\partial }{\partial p_{e}}\left(
L_{bd}\circ \varphi _{H}\right) \\
& -\left( L_{ce}\circ \varphi _{H}\right) \cdot \rho \overset{\ast }{V}%
_{d}^{ef}\cdot \left( L_{bf}\circ \varphi _{H}\right) .%
\end{array}%
\leqno(6.6)^{\prime }
\end{equation*}

\emph{In the classical case, }$\left( \rho ,\eta ,h\right) =\left(
Id_{TM},Id_{M},Id_{M}\right) $\emph{, we obtain}%
\begin{equation*}
\begin{array}{cc}
H_{jk}^{i}\circ \varphi _{H} & =\overset{\ast }{H}_{jk}^{i},%
\end{array}%
\leqno(6.3)^{\prime \prime }
\end{equation*}%
\begin{equation*}
\begin{array}{rl}
\left( H_{jk}^{i}\cdot \frac{\partial ^{2}L}{\partial y^{i}\partial y^{h}}%
\right) \circ \varphi _{H} & =\frac{\partial }{\partial x^{k}}\left( \frac{%
\partial ^{2}L}{\partial y^{j}\partial y^{h}}\circ \varphi _{H}\right) \\
& +\Gamma _{jk}\cdot \frac{\partial }{\partial p_{e}}\left( \frac{\partial
^{2}L}{\partial y^{e}\partial y^{h}}\circ \varphi _{H}\right) \\
& -\overset{\ast }{H}_{jk}^{i}\cdot \left( \frac{\partial ^{2}L}{\partial
y^{i}\partial y^{h}}\circ \varphi _{H}\right) ,%
\end{array}%
\leqno(6.4)^{\prime \prime }
\end{equation*}%
\begin{equation*}
\begin{array}{cc}
V_{jk}^{i}\circ \varphi _{H} & =\rho \overset{\ast }{V}_{j}^{ih}\cdot \left(
\frac{\partial ^{2}L}{\partial y^{h}\partial y^{k}}\circ \varphi _{H}\right)%
\end{array}%
\leqno(6.5)^{\prime \prime }
\end{equation*}%
\emph{and}%
\begin{equation*}
\begin{array}{rl}
\left( V_{jk}^{i}\cdot \frac{\partial ^{2}L}{\partial y^{i}\partial y^{h}}%
\right) \circ \varphi _{H} & =\left( \frac{\partial ^{2}L}{\partial
y^{k}\partial y^{e}}\circ \varphi _{H}\right) \cdot \frac{\partial }{%
\partial p_{e}}\left( \frac{\partial ^{2}L}{\partial y^{j}\partial y^{h}}%
\circ \varphi _{H}\right) \\
& -\left( \frac{\partial ^{2}L}{\partial y^{k}\partial y^{e}}\circ \varphi
_{H}\right) \cdot \rho \overset{\ast }{V}_{h}^{ef}\cdot \left( \frac{%
\partial ^{2}L}{\partial y^{j}\partial y^{f}}\circ \varphi _{H}\right) .%
\end{array}%
\leqno(6.6)^{\prime \prime }
\end{equation*}

\textbf{Theorem 6.2 }\emph{Dual, if}
\begin{equation*}
\begin{array}{c}
\left( E,\pi ,M\right) \overset{\mathcal{HL}}{\widetilde{_{\left( \rho ,\eta
,h\right) }}}\left( \overset{\ast }{E},\overset{\ast }{\pi },M\right)%
\end{array}%
\end{equation*}%
\emph{and }%
\begin{equation*}
\begin{array}{c}
\Gamma \left( \left( \rho ,\eta \right) T\varphi _{H},\varphi _{H}\right)
\left( \left( \rho ,\eta \right) \overset{\ast }{D}_{X}Y\right) =\left( \rho
,\eta \right) D_{\Gamma \left( \left( \rho ,\eta \right) T\varphi
_{H},\varphi _{H}\right) X}\Gamma \left( \left( \rho ,\eta \right) T\varphi
_{H},\varphi _{H}\right) Y,%
\end{array}%
\end{equation*}%
\emph{for any }$X,Y\in \Gamma \left( \left( \rho ,\eta \right) T\overset{%
\ast }{E},\left( \rho ,\eta \right) \tau _{\overset{\ast }{E}},\overset{\ast
}{E}\right) $\emph{, then we obtain}%
\begin{equation*}
\begin{array}{cc}
\left( \rho ,\eta \right) \overset{\ast }{H}_{\beta \gamma }^{\alpha }\circ
\varphi _{L} & =\left( \rho ,\eta \right) H_{\beta \gamma }^{\alpha },%
\end{array}%
\leqno(6.7)
\end{equation*}%
\begin{equation*}
\begin{array}{rl}
\left( \left( \rho ,\eta \right) \overset{\ast }{H}_{b\gamma }^{a}\cdot
H^{bc}\right) \circ \varphi _{L} & =\left( \rho _{\gamma }^{k}{\circ }h{%
\circ }\pi \right) \cdot \frac{\partial }{\partial x^{k}}\left( H^{ac}\circ
\varphi _{L}\right) \\
& +\left( \rho ,\eta \right) \Gamma _{\gamma }^{b}\cdot \frac{\partial }{%
\partial y^{b}}\left( H^{ac}\circ \varphi _{L}\right) \\
& -\left( \rho ,\eta \right) H_{b\gamma }^{a}\cdot \left( H^{bc}\circ
\varphi _{L}\right) ,%
\end{array}%
\leqno(6.8)
\end{equation*}%
\begin{equation*}
\begin{array}{cc}
\left( \rho ,\eta \right) \overset{\ast }{V}_{\beta }^{\alpha c}\circ
\varphi _{L} & =\left( \rho ,\eta \right) V_{\beta c}^{\alpha }\cdot \left(
H^{cd}\circ \varphi _{L}\right)%
\end{array}%
\leqno(6.9)
\end{equation*}%
\emph{and}%
\begin{equation*}
\begin{array}{rl}
\left( \left( \rho ,\eta \right) \overset{\ast }{V}_{a}^{bc}\cdot
H^{ad}\right) \circ \varphi _{L} & =\left( H^{ce}\circ \varphi _{H}\right)
\cdot \frac{\partial }{\partial y^{e}}\left( H^{bd}\circ \varphi _{L}\right)
\\
& -\left( H^{ce}\circ \varphi _{L}\right) \cdot \left( \rho ,\eta \right)
V_{ef}^{d}\cdot \left( H^{bf}\circ \varphi _{L}\right) .%
\end{array}%
\leqno(6.10)
\end{equation*}

\textbf{Corollary 6.1 }\emph{In the particular case of Lie algebroids,} $%
\left( \eta ,h\right) =\left( Id_{M},Id_{M}\right) $\emph{, we obtain }%
\begin{equation*}
\begin{array}{cc}
\rho \overset{\ast }{H}_{\beta \gamma }^{\alpha }\circ \varphi _{L} & =\rho
H_{\beta \gamma }^{\alpha },%
\end{array}%
\leqno(6.7)^{\prime }
\end{equation*}%
\begin{equation*}
\begin{array}{rl}
\left( \rho \overset{\ast }{H}_{b\gamma }^{a}\cdot H^{bc}\right) \circ
\varphi _{L} & =\left( \rho _{\gamma }^{k}{\circ }\pi \right) \cdot \frac{%
\partial }{\partial x^{k}}\left( H^{ac}\circ \varphi _{L}\right) \\
& +\rho \Gamma _{\gamma }^{b}\cdot \frac{\partial }{\partial y^{b}}\left(
H^{ac}\circ \varphi _{L}\right) \\
& -\rho H_{b\gamma }^{a}\cdot \left( H^{bc}\circ \varphi _{L}\right) ,%
\end{array}%
\leqno(6.8)^{\prime }
\end{equation*}%
\begin{equation*}
\begin{array}{cc}
\rho \overset{\ast }{V}_{\beta }^{\alpha c}\circ \varphi _{L} & =\rho
V_{\beta c}^{\alpha }\cdot \left( H^{cd}\circ \varphi _{L}\right)%
\end{array}%
\leqno(6.9)^{\prime }
\end{equation*}%
\emph{and}%
\begin{equation*}
\begin{array}{rl}
\left( \rho \overset{\ast }{V}_{a}^{bc}\cdot H^{ad}\right) \circ \varphi _{L}
& =\left( H^{ce}\circ \varphi _{H}\right) \cdot \frac{\partial }{\partial
y^{e}}\left( H^{bd}\circ \varphi _{L}\right) \\
& -\left( H^{ce}\circ \varphi _{L}\right) \cdot \left( \rho ,\eta \right)
V_{ef}^{d}\cdot \left( H^{bf}\circ \varphi _{L}\right) .%
\end{array}%
\leqno(6.10)^{\prime }
\end{equation*}

\emph{In the classical case, }$\left( \rho ,\eta ,h\right) =\left(
Id_{TM},Id_{M},Id_{M}\right) $\emph{, we obtain}%
\begin{equation*}
\begin{array}{cc}
\overset{\ast }{H}_{jk}^{i}\circ \varphi _{L} & =H_{jk}^{i},%
\end{array}%
\leqno(6.7)^{\prime }
\end{equation*}%
\begin{equation*}
\begin{array}{rl}
\left( \overset{\ast }{H}_{jk}^{i}\cdot \frac{\partial ^{2}H}{\partial
p_{j}\partial p_{h}}\right) \circ \varphi _{L} & =\frac{\partial }{\partial
x^{k}}\left( \frac{\partial ^{2}H}{\partial p_{i}\partial p_{h}}\circ
\varphi _{L}\right) \\
& +\Gamma _{k}^{e}\cdot \frac{\partial }{\partial y^{e}}\left( \frac{%
\partial ^{2}H}{\partial p_{i}\partial p_{h}}\circ \varphi _{L}\right) \\
& -\rho H_{jk}^{i}\cdot \left( \frac{\partial ^{2}H}{\partial p_{j}\partial
p_{h}}\circ \varphi _{L}\right) ,%
\end{array}%
\leqno(6.8)^{\prime }
\end{equation*}%
\begin{equation*}
\begin{array}{cc}
\overset{\ast }{V}_{j}^{ik}\circ \varphi _{L} & =\rho V_{jh}^{i}\cdot \left(
\frac{\partial ^{2}H}{\partial p_{h}\partial p_{k}}\circ \varphi _{L}\right)%
\end{array}%
\leqno(6.9)^{\prime }
\end{equation*}%
\emph{and}%
\begin{equation*}
\begin{array}{rl}
\left( \rho \overset{\ast }{V}_{i}^{jk}\cdot \frac{\partial ^{2}H}{\partial
p_{i}\partial p_{h}}\right) \circ \varphi _{L} & =\left( \frac{\partial ^{2}H%
}{\partial p_{k}\partial p_{e}}\circ \varphi _{H}\right) \cdot \frac{%
\partial }{\partial y^{e}}\left( \frac{\partial ^{2}H}{\partial
p_{j}\partial p_{h}}\circ \varphi _{L}\right) \\
& -\left( \frac{\partial ^{2}H}{\partial p_{k}\partial p_{e}}\circ \varphi
_{L}\right) \cdot V_{ef}^{h}\cdot \left( \frac{\partial ^{2}H}{\partial
p_{j}\partial p_{f}}\circ \varphi _{L}\right) .%
\end{array}%
\leqno(6.10)^{\prime }
\end{equation*}

\section{Duality between mechanical $\left( \protect\rho ,\protect\eta %
\right) $-systems}

Let $(\left( E,\pi ,M\right) ,F_{e},(\rho ,\eta )\Gamma )$ be a mechanical $%
\left( \rho ,\eta \right) $-system.

If $g\in \mathbf{Man}\left( E,E\right) $ such that $\left( g,h\right) $ is a
$\mathbf{B}^{\mathbf{v}}$\textit{-}morphism locally invertible of $\left(
E,\pi ,M\right) $ source and $\left( E,\pi ,M\right) $ target, on components
$g_{b}^{a}$, then the $\mathbf{Mod}$-endomorphism
\begin{equation*}
\begin{array}{rcl}
\Gamma \left( \left( \rho ,\eta \right) TE,\left( \rho ,\eta \right) \tau
_{E},E\right) & ^{\underrightarrow{\mathcal{J}_{\left( g,h\right) }}} &
\Gamma \left( \left( \rho ,\eta \right) TE,\left( \rho ,\eta \right) \tau
_{E},E\right) \\
\tilde{Z}^{a}\tilde{\partial}_{a}+Y^{b}\overset{\cdot }{\tilde{\partial}}_{b}
& \longmapsto & \left( \tilde{g}_{a}^{b}\circ h\circ \pi \right) \tilde{Z}%
^{a}\overset{\cdot }{\tilde{\partial}}_{b}%
\end{array}%
\leqno(7.1)
\end{equation*}%
is the almost tangent structure associated to the\emph{\ }$\mathbf{B}^{%
\mathbf{v}}$-morphism $\left( g,h\right) $.

The\textit{\ }vertical section
\begin{equation*}
\begin{array}{c}
\mathbb{C}\mathbf{=}y^{a}\overset{\cdot }{\tilde{\partial}}_{a}%
\end{array}%
\leqno(7.2)
\end{equation*}%
is the Liouville section.

Let $\left( \left( \overset{\ast }{E},\overset{\ast }{\pi },M\right) ,%
\overset{\ast }{F}_{e},(\rho ,\eta )\Gamma \right) $ be a dual mechanical $%
\left( \rho ,\eta \right) $-system.

If $g\in \mathbf{Man}\left( \overset{\ast }{E},E\right) $ be such that $%
\left( g,h\right) $ is a $\mathbf{B}^{\mathbf{v}}$\textit{-}morphism locally
invertible of $\left( \overset{\ast }{E},\overset{\ast }{\pi },M\right) $
source and $\left( E,\pi ,M\right) $ target, on components $g^{ab}$, then
the $\mathbf{Mod}$-endomorphism
\begin{equation*}
\begin{array}{rcl}
\Gamma \left( \left( \rho ,\eta \right) T\overset{\ast }{E},\left( \rho
,\eta \right) \tau _{\overset{\ast }{E}},\overset{\ast }{E}\right) & ^{%
\underrightarrow{\overset{\ast }{\mathcal{J}}_{\left( g,h\right) }}} &
\Gamma \left( \left( \rho ,\eta \right) T\overset{\ast }{E},\left( \rho
,\eta \right) \tau _{\overset{\ast }{E}},\overset{\ast }{E}\right) \\
\tilde{Z}^{a}\overset{\ast }{\tilde{\partial}}_{a}+Y_{b}\overset{\cdot }{%
\tilde{\partial}}^{b} & \longmapsto & \left( \tilde{g}_{ba}\circ h\circ
\overset{\ast }{\pi }\right) \tilde{Z}^{a}\overset{\cdot }{\tilde{\partial}}%
^{b}%
\end{array}%
\leqno(7.1)^{\prime }
\end{equation*}%
is the almost tangent structure associated to the $\mathbf{B}^{\mathbf{v}}$%
-morphism $\left( g,h\right) $.

The vertical section
\begin{equation*}
\begin{array}{c}
\overset{\ast }{\mathbb{C}}\mathbf{=}p_{b}\overset{\cdot }{\tilde{\partial}}%
^{b}%
\end{array}%
\leqno(7.2)^{\prime }
\end{equation*}%
is the Liouville section.

Let
\begin{equation*}
\begin{array}{l}
S=y^{b}\left( g_{b}^{a}\circ h\circ \pi \right) \tilde{\partial}_{a}-2\left(
G^{a}-\frac{1}{4}F^{a}\right) \overset{\cdot }{\tilde{\partial}}_{a}%
\end{array}%
\leqno(7.3)
\end{equation*}%
be the $\left( \rho ,\eta \right) $-semispray associated to mechanical $%
\left( \rho ,\eta \right) $-system $\left( \left( E,\pi ,M\right)
,F_{e},\left( \rho ,\eta \right) \Gamma \right) $\ and from locally
invertible $\mathbf{B}^{\mathbf{v}}$-morphism $\left( g,h\right) $ and let
\begin{equation*}
\begin{array}{l}
\overset{\ast }{S}=p_{b}\left( g^{ab}\circ h\circ \overset{\ast }{\pi }%
\right) \overset{\ast }{\tilde{\partial}}_{a}-2\left( G_{a}-\frac{1}{4}%
F_{a}\right) \overset{\cdot }{\tilde{\partial}}^{a}%
\end{array}%
\leqno(7.3)^{\prime }
\end{equation*}%
be the $\left( \rho ,\eta \right) $-semispray associated to the mechanical $%
\left( \rho ,\eta \right) $-system $\left( \left( \overset{\ast }{E},\overset%
{\ast }{\pi },M\right) ,\overset{\ast }{F}_{e},\left( \rho ,\eta \right)
\Gamma \right) $\ and from locally invertible $\mathbf{B}^{\mathbf{v}}$%
-morphism $\left( g,h\right) .$

\medskip \textbf{Theorem 7.1 }\emph{\ If }$\Gamma \left( \left( \rho ,\eta
\right) T\varphi _{L},\varphi _{L}\right) \left( S\right) =\overset{\ast }{S}%
,$ \emph{then we obtain:}%
\begin{equation*}
\begin{array}{cc}
y^{b}\left( g_{b}^{a}\circ h\circ \pi \right) \circ \varphi _{H} &
=p_{b}\left( g^{ab}\circ h\circ \overset{\ast }{\pi }\right)%
\end{array}%
\leqno(7.4)
\end{equation*}%
\emph{and }%
\begin{equation*}
\begin{array}{cl}
2\left( G_{b}-\frac{1}{4}F_{b}\right) & =2\left[ \left( G^{a}-\frac{1}{4}%
F^{a}\right) L_{ab}\right] \circ \varphi _{H} \\
& -y^{c}\left\{ \left[ \left( g_{c}^{a}\rho _{a}^{i}\right) \circ h\circ \pi %
\right] L_{ib}\right\} \circ \varphi _{H}.%
\end{array}%
\leqno(7.5)
\end{equation*}

\textbf{Corollary 7.1 }\emph{In the particular case of Lie algebroids,} $%
\left( \eta ,h\right) =\left( Id_{M},Id_{M}\right) $\emph{, we obtain}%
\begin{equation*}
\begin{array}{cl}
2\left( G_{b}-\frac{1}{4}F_{b}\right) & =2\left[ \left( G^{a}-\frac{1}{4}%
F^{a}\right) L_{ab}\right] \circ \varphi _{H} \\
& -y^{c}\left\{ \left[ \left( g_{c}^{a}\rho _{a}^{i}\right) \circ \pi \right]
L_{ib}\right\} \circ \varphi _{H}.%
\end{array}%
\leqno(7.4)^{\prime }
\end{equation*}

\emph{In the classical case, }$\left( \rho ,\eta ,h\right) =\left(
Id_{TM},Id_{M},Id_{M}\right) $\emph{, we obtain the equality} \emph{implies
the equality}%
\begin{equation*}
\begin{array}{cl}
2\left( G_{i}-\frac{1}{4}F_{i}\right) & =2\left[ \left( G^{i}-\frac{1}{4}%
F^{i}\right) \frac{\partial ^{2}L}{\partial y^{i}\partial y^{j}}\right]
\circ \varphi _{H} \\
& -y^{i}\frac{\partial ^{2}L}{\partial x^{i}\partial y^{j}}\circ \varphi
_{H}.%
\end{array}%
\leqno(7.5)^{\prime \prime }
\end{equation*}

\textbf{Theorem 7.2 }\emph{\ Dual, if }$\Gamma \left( \left( \rho ,\eta
\right) T\varphi _{H},\varphi _{H}\right) \left( \overset{\ast }{S}\right)
=S,$ \emph{then we obtain:}%
\begin{equation*}
\begin{array}{cc}
p_{b}\left( g^{ba}\circ h\circ \overset{\ast }{\pi }\right) \circ \varphi
_{L} & =y^{b}\left( g_{b}^{a}\circ h\circ \pi \right)%
\end{array}%
\leqno(7.6)
\end{equation*}%
\emph{and }%
\begin{equation*}
\begin{array}{cl}
2\left( G^{a}-\frac{1}{4}F^{a}\right) & =2\left[ \left( G_{b}-\frac{1}{4}%
F_{b}\right) H^{ab}\right] \circ \varphi _{L} \\
& -p_{c}\left\{ \left[ \left( g^{ac}\rho _{a}^{i}\right) \circ h\circ
\overset{\ast }{\pi }\right] H_{i}^{b}\right\} \circ \varphi _{L}.%
\end{array}%
\leqno(7.7)
\end{equation*}

\textbf{Corollary 7.2 }\emph{In the particular case of Lie algebroids,} $%
\left( \eta ,h\right) =\left( Id_{M},Id_{M}\right) $\emph{, we obtain}%
\begin{equation*}
\begin{array}{cl}
2\left( G^{a}-\frac{1}{4}F^{a}\right) & =2\left[ \left( G_{b}-\frac{1}{4}%
F_{b}\right) H^{ab}\right] \circ \varphi _{L} \\
& -p_{c}\left\{ \left[ \left( g^{ac}\rho _{a}^{i}\right) \circ \overset{\ast
}{\pi }\right] H_{i}^{b}\right\} \circ \varphi _{L}.%
\end{array}%
\leqno(7.7)^{\prime }
\end{equation*}

\section{Duality between Lagrange and Hamilton mechanical $\left( \protect%
\rho ,\protect\eta \right) $-systems}

Let $\left( \left( E,\pi ,M\right) ,F_{e},L\right) $ be an arbitrarily
Lagrange mechanical $\left( \rho ,\eta \right) $-system.

Let $d^{\left( \rho ,\eta \right) TE}$ be the exterior differentiation
operator associated to the exterior differential $\mathcal{F}\left( E\right)
$-algebra
\begin{equation*}
\left( \Lambda \left( \left( \rho ,\eta \right) TE,\left( \rho ,\eta \right)
\tau _{E},E\right) ,+,\cdot ,\wedge \right)
\end{equation*}%
and let $\left( g,h\right) $\ be a locally invertible $\mathbf{B}^{\mathbf{v}%
}$-morphism of $\left( E,\pi ,M\right) $ source and $\left( E,\pi ,M\right) $
target.

Let $\left( \left( \overset{\ast }{E},\overset{\ast }{\pi },M\right) ,%
\overset{\ast }{F}_{e},H\right) $ be an Hamilton mechanical $\left( \rho
,\eta \right) $-system, where the regular Hamiltonian $H$ is the Legendre
transformation of the regular Lagrangian $L.$

Let $d^{\left( \rho ,\eta \right) T\overset{\ast }{E}}$ be the exterior
differentiation operator associated to the exterior differential $\mathcal{F}%
\left( \overset{\ast }{E}\right) $-algebra
\begin{equation*}
\left( \Lambda \left( \left( \rho ,\eta \right) T\overset{\ast }{E},\left(
\rho ,\eta \right) \tau _{\overset{\ast }{E}},\overset{\ast }{E}\right)
,+,\cdot ,\wedge \right) .
\end{equation*}%
and let $\left( g,h\right) $\ be a locally invertible $\mathbf{B}^{\mathbf{v}%
}$-morphism of $\left( \overset{\ast }{E},\overset{\ast }{\pi },M\right) $
source and $\left( E,\pi ,M\right) $ target.

The $1$-form
\begin{equation*}
\begin{array}{c}
\theta _{L}=\left( \tilde{g}_{a}^{e}\circ h\circ \pi \cdot L_{e}\right) d%
\tilde{z}^{a}%
\end{array}%
\leqno(8.1)
\end{equation*}%
is called the $1$\emph{-form of Poincar\'{e}-Cartan type associated to the
Lagrangian }$L$ \emph{and to the locally invertible }$\mathbf{B}^{\mathbf{v}%
} $\emph{-morphism }$\left( g,h\right) $.\medskip\ We obtain easily:
\begin{equation*}
\begin{array}[t]{l}
\theta _{L}\left( \tilde{\partial}_{a}\right) =\left( \tilde{g}_{b}^{e}\circ
h\circ \pi \right) L_{e},\,\,\ \theta _{L}\left( \overset{\cdot }{\tilde{%
\partial}}_{b}\right) =0.%
\end{array}%
\leqno(8.2)
\end{equation*}

The $1$-form
\begin{equation*}
\begin{array}{c}
\theta _{H}=\left( \tilde{g}_{ae}\circ h\circ \overset{\ast }{\pi }\right)
H^{e}d\tilde{z}^{a}%
\end{array}%
\leqno(8.1)^{\prime }
\end{equation*}%
will be called the $1$\emph{-form of Poincar\'{e}-Cartan type associated to
the regular Hamiltonian }$H$ \emph{and from locally invertible }$\mathbf{B}^{%
\mathbf{v}}$\emph{-morphism }$\left( g,h\right) $.\medskip\ We obtain
easily:
\begin{equation*}
\begin{array}[t]{l}
\theta _{H}\left( \overset{\ast }{\tilde{\partial}}_{b}\right) =\left(
\tilde{g}_{be}\circ h\circ \overset{\ast }{\pi }\right) H^{e},\,\,\ \theta
_{H}\left( \overset{\cdot }{\tilde{\partial}}^{b}\right) =0.%
\end{array}%
\leqno(8.2)^{\prime }
\end{equation*}

\textbf{Theorem 8.1 }If $\left( \left( \rho ,\eta \right) T\varphi
_{L},\varphi _{L}\right) ^{\ast }\left( \theta _{L}\right) =\theta _{H},$
then
\begin{equation*}
\begin{array}[b]{c}
\left( \tilde{g}_{ae}\circ h\circ \overset{\ast }{\pi }\right) H^{e}\circ
\varphi _{L}=\left( \tilde{g}_{b}^{e}\circ h\circ \pi \right) L_{e}%
\end{array}%
\leqno(8.3)
\end{equation*}

Dual, if $\left( \left( \rho ,\eta \right) T\varphi _{H},\varphi _{H}\right)
^{\ast }\left( \theta _{H}\right) =\theta _{L},$ then
\begin{equation*}
\begin{array}[b]{c}
\left( \tilde{g}_{b}^{e}\circ h\circ \pi \right) L_{e}\circ \varphi
_{H}=\left( \tilde{g}_{ae}\circ h\circ \overset{\ast }{\pi }\right) H^{e}.%
\end{array}%
\leqno(8.3)^{\prime }
\end{equation*}

The $2$-form
\begin{equation*}
\begin{array}[b]{c}
\omega _{L}=d^{\left( \rho ,\eta \right) TE}\theta _{L}%
\end{array}%
\leqno(8.4)
\end{equation*}%
is called the $2$\emph{-form of Poincar\'{e}-Cartan type associated to the
Lagrangian }$L$\emph{\ and to the locally invertible }$\mathbf{B}^{\mathbf{v}%
}$\emph{-morphism }$\left( g,h\right) $.\medskip\ By the definition of $%
d^{\left( \rho ,\eta \right) TE},$ we obtain:
\begin{equation*}
\begin{array}{ll}
\omega _{L}\left( U,V\right) & \displaystyle=\Gamma \left( \tilde{\rho}%
,Id_{E}\right) \left( U\right) \left( \theta _{L}\left( V\right) \right)
\vspace*{2mm} \\
& \displaystyle-\,\Gamma \left( \tilde{\rho},Id_{E}\right) \left( V\right)
\left( \theta _{L}\left( U\right) \right) -\theta _{L}\left( \left[ U,V%
\right] _{\left( \rho ,\eta \right) TE}\right) ,%
\end{array}%
\leqno(8.5)
\end{equation*}%
for any $U,V\in \Gamma \left( \left( \rho ,\eta \right) TE,\left( \rho ,\eta
\right) \tau _{E},E\right) $.\smallskip

The $2$-form
\begin{equation*}
\begin{array}[b]{c}
\omega _{H}=d^{\left( \rho ,\eta \right) T\overset{\ast }{E}}\theta _{H}%
\end{array}%
\leqno(8.4)^{\prime }
\end{equation*}%
will be called the $2$\emph{-form of Poincar\'{e}-Cartan type associated to
the Hamiltonian }$H$\emph{\ and to the locally invertible }$\mathbf{B}^{%
\mathbf{v}}$\emph{-morphism }$\left( g,h\right) $.\medskip\ By the
definition of $d^{\left( \rho ,\eta \right) T\overset{\ast }{E}},$ we
obtain:
\begin{equation*}
\begin{array}{ll}
\omega _{H}\left( U,V\right) & \displaystyle=\Gamma \left( \overset{\ast }{%
\tilde{\rho}},Id_{\overset{\ast }{E}}\right) \left( U\right) \left( \theta
_{H}\left( V\right) \right) \vspace*{2mm} \\
& \displaystyle-\,\Gamma \left( \overset{\ast }{\tilde{\rho}},Id_{\overset{%
\ast }{E}}\right) \left( V\right) \left( \theta _{H}\left( U\right) \right)
-\theta _{H}\left( \left[ U,V\right] _{\left( \rho ,\eta \right) T\overset{%
\ast }{E}}\right) ,%
\end{array}%
\leqno(8.5)^{\prime }
\end{equation*}%
for any $U,V\in \Gamma \left( \left( \rho ,\eta \right) T\overset{\ast }{E}%
,\left( \rho ,\eta \right) \tau _{\overset{\ast }{E}},\overset{\ast }{E}%
\right) $.\smallskip

\textbf{Theorem 8.2 }\emph{If }$\left( \left( \rho ,\eta \right) T\varphi
_{L},\varphi _{L}\right) ^{\ast }\left( \omega _{L}\right) =\omega _{H},$
\emph{then}%
\begin{equation*}
(8.6)%
\begin{array}{l}
\left[ \left( \rho _{a}^{i}\circ h\circ \overset{\ast }{\pi }\right) \frac{%
\partial \left( \left( \tilde{g}_{be}\circ h\circ \overset{\ast }{\pi }%
\right) H^{e}\right) }{\partial x^{i}}\right] \circ \varphi _{L}-\left[
\left( \rho _{b}^{j}\circ h\circ \overset{\ast }{\pi }\right) \frac{\partial
\left( \left( \tilde{g}_{ae}\circ h\circ \overset{\ast }{\pi }\right)
H^{e}\right) }{\partial x^{j}}\right] \circ \varphi _{L} \\
\left[ \left( L_{ab}^{c}\circ h\circ \overset{\ast }{\pi }\right) \left(
\tilde{g}_{ce}\circ h\circ \overset{\ast }{\pi }\right) H^{e}\right] \circ
\varphi _{L}-\left[ \left( \left( \rho _{b}^{j}\circ h\circ \overset{\ast }{%
\pi }\right) L_{jd}\circ \varphi _{H}\right) \frac{\partial \left( \left(
\tilde{g}_{be}\circ h\circ \overset{\ast }{\pi }\right) H^{e}\right) }{%
\partial p_{d}}\right] \circ \varphi _{L} \\
-\left[ \left( \left( \rho _{a}^{i}\circ h\circ \overset{\ast }{\pi }\right)
L_{id}\right) \circ \varphi _{H}\frac{\partial \left( \left( \tilde{g}%
_{ae}\circ h\circ \overset{\ast }{\pi }\right) H^{e}\right) }{\partial p_{d}}%
\right] \circ \varphi _{L} \\
=\left( \rho _{a}^{i}\circ h\circ \pi \right) \frac{\partial \left( \left(
\tilde{g}_{b}^{e}\circ h\circ \pi \right) L_{e}\right) }{\partial x^{i}}%
-\left( \rho _{b}^{j}\circ h\circ \pi \right) \frac{\partial \left( \left(
\tilde{g}_{a}^{e}\circ h\circ \pi \right) L_{e}\right) }{\partial x^{j}}%
-\left( L_{ab}^{c}\circ h\circ \pi \right) \left( \tilde{g}_{c}^{e}\circ
h\circ \pi \right) L_{e}%
\end{array}%
\end{equation*}%
\emph{and}
\begin{equation*}
\begin{array}[b]{c}
\left[ L_{bc}\circ \varphi _{H}\frac{\partial \left( \left( \tilde{g}%
_{ae}\circ h\circ \overset{\ast }{\pi }\right) H^{e}\right) }{\partial p_{c}}%
\right] \circ \varphi _{L}=\frac{\partial \left( \left( \tilde{g}%
_{a}^{e}\circ h\circ \pi \right) L_{e}\right) }{\partial y^{b}}.%
\end{array}%
\leqno(8.7)
\end{equation*}

\textbf{Corollary 8.1 }\emph{In the particular case of Lie algebroids, }$%
\left( \eta ,h\right) =\left( Id_{M},Id_{M}\right) $\emph{, we obtain}
\begin{equation*}
\begin{array}{c}
\left[ \left( \rho _{a}^{i}\circ \overset{\ast }{\pi }\right) \frac{\partial
\left( \left( \tilde{g}_{be}\circ \overset{\ast }{\pi }\right) H^{e}\right)
}{\partial x^{i}}\right] \circ \varphi _{L}-\left[ \left( \rho _{b}^{j}\circ
\overset{\ast }{\pi }\right) \frac{\partial \left( \left( \tilde{g}%
_{ae}\circ \overset{\ast }{\pi }\right) H^{e}\right) }{\partial x^{j}}\right]
\circ \varphi _{L} \\
\left[ \left( L_{ab}^{c}\circ \overset{\ast }{\pi }\right) \left( \tilde{g}%
_{ce}\circ \overset{\ast }{\pi }\right) H^{e}\right] \circ \varphi _{L}-%
\left[ \left( \left( \rho _{b}^{j}\circ \overset{\ast }{\pi }\right)
L_{jd}\circ \varphi _{H}\right) \frac{\partial \left( \left( \tilde{g}%
_{be}\circ \overset{\ast }{\pi }\right) H^{e}\right) }{\partial p_{d}}\right]
\circ \varphi _{L} \\
-\left[ \left( \left( \rho _{a}^{i}\circ \overset{\ast }{\pi }\right)
L_{id}\right) \circ \varphi _{H}\frac{\partial \left( \left( \tilde{g}%
_{ae}\circ \overset{\ast }{\pi }\right) H^{e}\right) }{\partial p_{d}}\right]
\circ \varphi _{L} \\
=\left( \rho _{a}^{i}\circ \pi \right) \frac{\partial \left( \left( \tilde{g}%
_{b}^{e}\circ \pi \right) L_{e}\right) }{\partial x^{i}}-\left( \rho
_{b}^{j}\circ \pi \right) \frac{\partial \left( \left( \tilde{g}%
_{a}^{e}\circ \pi \right) L_{e}\right) }{\partial x^{j}}-\left(
L_{ab}^{c}\circ \pi \right) \left( \tilde{g}_{c}^{e}\circ \pi \right) L_{e}%
\end{array}%
\leqno(8.6)^{\prime }
\end{equation*}%
\emph{and}
\begin{equation*}
\begin{array}[b]{c}
\left[ L_{bc}\circ \varphi _{H}\frac{\partial \left( \left( \tilde{g}%
_{ae}\circ \overset{\ast }{\pi }\right) H^{e}\right) }{\partial p_{c}}\right]
\circ \varphi _{L}=\frac{\partial \left( \left( \tilde{g}_{a}^{e}\circ \pi
\right) L_{e}\right) }{\partial y^{b}}.%
\end{array}%
\leqno(8.7)^{\prime }
\end{equation*}

\textbf{Theorem 8.3 }\emph{If} $\left( \left( \rho ,\eta \right) T\varphi
_{H},\varphi _{H}\right) ^{\ast }\left( \omega _{H}\right) =\omega _{L},$
\emph{then}%
\begin{equation*}
(8.8)%
\begin{array}{l}
\left[ \left( \rho _{a}^{i}\circ h\circ \pi \right) \frac{\partial \left(
\left( \tilde{g}_{b}^{e}\circ h\circ \pi \right) L_{e}\right) }{\partial
x^{i}}\right] \circ \varphi _{H}-\left[ \left( \rho _{b}^{j}\circ h\circ \pi
\right) \frac{\partial \left( \left( \tilde{g}_{a}^{e}\circ h\circ \pi
\right) L_{e}\right) }{\partial x^{j}}\right] \circ \varphi _{H} \\
\left[ \left( L_{ab}^{c}\circ h\circ \pi \right) \left( \tilde{g}%
_{c}^{e}\circ h\circ \pi \right) L_{e}\right] \circ \varphi _{H}-\left[
\left( \left( \rho _{b}^{j}\circ h\circ \pi \right) H_{j}^{d}\circ \varphi
_{L}\right) \frac{\partial \left( \left( \tilde{g}_{b}^{e}\circ h\circ \pi
\right) L_{e}\right) }{\partial y^{d}}\right] \circ \varphi _{H} \\
-\left[ \left( \left( \rho _{a}^{i}\circ h\circ \pi \right) H_{i}^{d}\circ
\varphi _{L}\right) \frac{\partial \left( \left( \tilde{g}_{a}^{e}\circ
h\circ \pi \right) L_{e}\right) }{\partial y^{d}}\right] \circ \varphi _{H}
\\
=\left( \rho _{a}^{i}\circ h\circ \overset{\ast }{\pi }\right) \frac{%
\partial \left( \left( \tilde{g}_{be}\circ h\circ \overset{\ast }{\pi }%
\right) H^{e}\right) }{\partial x^{i}}-\left( \rho _{b}^{j}\circ h\circ
\overset{\ast }{\pi }\right) \frac{\partial \left( \left( \tilde{g}%
_{ae}\circ h\circ \overset{\ast }{\pi }\right) H^{e}\right) }{\partial x^{j}}%
-\left( L_{ab}^{c}\circ h\circ \overset{\ast }{\pi }\right) \left( \tilde{g}%
_{ce}\circ h\circ \overset{\ast }{\pi }\right) H^{e}%
\end{array}%
\end{equation*}%
\emph{and}
\begin{equation*}
\begin{array}[b]{c}
\left[ H^{bc}\circ \varphi _{L}\frac{\partial \left( \left( \tilde{g}%
_{a}^{e}\circ h\circ \pi \right) L_{e}\right) }{\partial y^{c}}\right] \circ
\varphi _{H}=\frac{\partial \left( \left( \tilde{g}_{ae}\circ h\circ \overset%
{\ast }{\pi }\right) H^{e}\right) }{\partial p_{b}}.%
\end{array}%
\leqno(8.9)
\end{equation*}

\textbf{Corollary 8.2 }\emph{In the particular case of Lie algebroids, }$%
\left( \eta ,h\right) =\left( Id_{M},Id_{M}\right) $\emph{, we obtain}%
\begin{equation*}
\begin{array}{l}
\left[ \left( \rho _{a}^{i}\circ \pi \right) \frac{\partial \left( \left(
\tilde{g}_{b}^{e}\circ \pi \right) L_{e}\right) }{\partial x^{i}}\right]
\circ \varphi _{H}-\left[ \left( \rho _{b}^{j}\circ \pi \right) \frac{%
\partial \left( \left( \tilde{g}_{a}^{e}\circ \pi \right) L_{e}\right) }{%
\partial x^{j}}\right] \circ \varphi _{H} \\
\left[ \left( L_{ab}^{c}\circ \pi \right) \left( \tilde{g}_{c}^{e}\circ \pi
\right) L_{e}\right] \circ \varphi _{H}-\left[ \left( \left( \rho
_{b}^{j}\circ \pi \right) H_{j}^{d}\circ \varphi _{L}\right) \frac{\partial
\left( \left( \tilde{g}_{b}^{e}\circ \pi \right) L_{e}\right) }{\partial
y^{d}}\right] \circ \varphi _{H} \\
-\left[ \left( \left( \rho _{a}^{i}\circ \pi \right) H_{i}^{d}\circ \varphi
_{L}\right) \frac{\partial \left( \left( \tilde{g}_{a}^{e}\circ \pi \right)
L_{e}\right) }{\partial y^{d}}\right] \circ \varphi _{H} \\
=\left( \rho _{a}^{i}\circ \overset{\ast }{\pi }\right) \frac{\partial
\left( \left( \tilde{g}_{be}\circ \overset{\ast }{\pi }\right) H^{e}\right)
}{\partial x^{i}}-\left( \rho _{b}^{j}\circ \overset{\ast }{\pi }\right)
\frac{\partial \left( \left( \tilde{g}_{ae}\circ \overset{\ast }{\pi }%
\right) H^{e}\right) }{\partial x^{j}}-\left( L_{ab}^{c}\circ \overset{\ast }%
{\pi }\right) \left( \tilde{g}_{ce}\circ \overset{\ast }{\pi }\right) H^{e}%
\end{array}%
\leqno(8.8)^{\prime }
\end{equation*}%
\emph{and}
\begin{equation*}
\begin{array}[b]{c}
\left[ H^{bc}\circ \varphi _{L}\frac{\partial \left( \left( \tilde{g}%
_{a}^{e}\circ \pi \right) L_{e}\right) }{\partial y^{c}}\right] \circ
\varphi _{H}=\frac{\partial \left( \left( \tilde{g}_{ae}\circ \overset{\ast }%
{\pi }\right) H^{e}\right) }{\partial p_{b}}.%
\end{array}%
\leqno(8.9)^{\prime }
\end{equation*}

\bigskip

\hfill
\begin{tabular}{c}
SECONDARY SCHOOL \textquotedblleft CORNELIUS RADU\textquotedblright , \\
RADINESTI VILLAGE, 217196, GORJ COUNTY, ROMANIA \\
e-mail: c\_arcus@yahoo.com, c\_arcus@radinesti.ro%
\end{tabular}

\end{document}